\documentclass[fleqn,usenatbib]{mnras}

\usepackage{newtxtext,newtxmath}

\usepackage[T1]{fontenc}
\usepackage[english]{babel}

\usepackage{graphicx}	
\usepackage{amsmath}	
\usepackage{booktabs}
\usepackage{mleftright}
\usepackage{siunitx}
\usepackage[english=british]{csquotes}
\usepackage[capitalise, noabbrev]{cleveref}

\DeclareSIUnit{\Msun}{\ensuremath{\mathrm{M}_\odot}}
\DeclareSIUnit{\hHubble}{\ensuremath{\mathnormal{h}}}

\title[Halo mass functions in mixed cold-fuzzy dark matter models]{Halo mass functions in mixed cold and fuzzy dark matter models}

\author[S.\ C.\ Johnston et al.]
{Sarah C.\ Johnston,$^{1}$\thanks{E-mail: \texttt{wgfr58@durham.ac.uk}}
Simon May,$^{2}$
Tibor Dome,$^{3}$
Sownak Bose,$^{1}$
Alastair Basden,$^{1}$
Carlton Baugh,$^{1}$
\newauthor
Anastasia Fialkov,$^{3,4}$
Alex Tocher$^{3,4}$
\\
$^{1}$Institute for Computational Cosmology, Department of Physics, Durham University, South Road, Durham, DH1 3LE, UK\\
$^{2}$Fakultät für Physik, Universität Bielefeld, Universitätsstraße 25, 33615 Bielefeld, Germany\\
$^{3}$Institute of Astronomy, University of Cambridge, Madingley Road, Cambridge, CB3 0HA, UK\\
$^{4}$Kavli Institute for Cosmology, Madingley Road, Cambridge, CB3 0HA, UK
}

\date{Accepted XXX. Received YYY; in original form ZZZ}

\pubyear{2026}

\begin{document}
\label{firstpage}
\pagerange{\pageref{firstpage}--\pageref{lastpage}}
\maketitle

\begin{abstract}
We investigate the impact of mixed cold and fuzzy dark matter (referred to as MDM) cosmologies on the halo mass function (HMF) using numerical simulations performed with the \texttt{AxiREPO} framework. We consider models in which an ultralight axion-like component with mass $m =10^{-24.5}\,\mathrm{eV}$ constitutes a fraction $f \leq 0.3$ of the total dark matter. To enable consistent halo identification in mixed-species scenarios, we develop a grid-based halo-finding pipeline that combines the particle-based cold dark matter (CDM) and wave-like fuzzy dark matter (FDM) components into a unified density field. We find that FDM traces the large-scale CDM distribution while suppressing small-scale structure through wave interference effects, leading to a reduction in the abundance of low-mass haloes and modifying the HMF in a manner dependent on redshift and FDM fraction. Increasing the FDM fraction produces a systematic downward shift in the HMF and modifies its high-mass slope. Motivated by these trends, we introduce a phenomenological model that maps CDM HMFs to their MDM counterparts using a suppression function with parameters dependent on redshift and FDM fraction. This model reproduces the simulated HMFs within approximately 0.1 to 0.2 dex across the parameter space explored ($1 \leq z \leq 4$, $f \leq 0.3$). Our results provide a computationally efficient method for predicting structure formation in MDM cosmologies without requiring dedicated simulations for each parameter choice, and establish a framework for exploring the impact of MDM on cosmological structure formation.
\end{abstract}

\begin{keywords}
dark matter -- large-scale structure of the universe -- methods: numerical
\end{keywords}



\section{Introduction}
The nature of dark matter remains one of the central open questions in cosmology. Our current cosmological model, $\Lambda$CDM, is highly successful in describing the large-scale structure of the Universe \citep{2021Planck}. A central component of this model is cold dark matter (CDM), which makes up $\sim \SI{85}{\percent}$ of the total matter density. However, apparent discrepancies were historically identified in dark matter-only simulations, including the core–cusp problem (\citealt{Flores1994, Moore1994, deBlock2010}), the missing satellite problem (\citealt{Klypin1999, Moore1999}) and the too-big-to-fail problem (\citealt{BoylanKolchin2011, BoylanKolchin2012}). Subsequent work over the past decade has demonstrated that many of these tensions can be significantly alleviated when baryonic processes such as stellar feedback, gas dynamics, and reionisation are included, as well as through improved observational constraints (\citealt{Bullock2017, 2016Fattahi, Wetzel2016, 2016Read, 2020SantosSantos, Sales2022, Jahn2023, Jackson2024}). As a result, their status as fundamental challenges to $\Lambda$CDM is now less clear, although uncertainties in the modelling of small-scale structure persist \citep{Efstathiou2025}. In addition, despite extensive efforts, there has been no conclusive direct detection of CDM particles (\citealt{Bertone2018, Baudis2025}), motivating the exploration of alternative dark matter scenarios.

One such alternative is fuzzy dark matter (FDM), an ultralight bosonic model in which the dark matter consists of axion-like particles (ALPs), motivated by both field theory (e.\,g.\ \citealt{Kim2016}) and string theory (e.\,g.\ \citealt{Gendler2024}). If these particles have masses $\lesssim \SI{e-21}{\eV}$ \citep{Hu2000}, they exhibit wave-like behaviour on astrophysical scales, with de Broglie wavelengths comparable to galactic sizes. This wave nature leads to the formation of a solitonic core in halo centres, i.\,e.\ a stationary, minimum-energy solution to the Schrödinger--Poisson equations \citep{Hui2017}. FDM is typically modelled as a minimally coupled classical field characterised by its mass. The model's popularity arises from both its theoretical motivation in high-energy theories and its production of distinctive, scale-dependent effects on structure formation, including the suppression of power below a characteristic scale and coherent wave interference phenomena \citep{Hu2000, Cicoli2022}.

Extensive studies of cosmologies in which the dark matter is \SI{100}{\percent} FDM have shown that such models are strongly constrained using many different observables \citep[e.\,g.][]{Dalal2022, Powell2023, Lazare2024, May2025}. Wave interference suppresses the small-scale power spectrum and delays structure formation (e.\,g.\ \citealt{Marsh2014, Mocz2017, 2021May, Dome2023elongation, May2023}), in tension with Lyman-$\alpha$ forest constraints \citep{Rogers2021}. This delayed formation affects halo growth and galaxy evolution \citep{Mocz2020}, and the presence of solitonic cores modifies galaxy rotation curves in ways that are not fully consistent with observations \citep{Schive2014, Banares2024}. Cosmological constraints, such as from the Cosmic Microwave Background (CMB), further limit both the allowed particle mass and fractional abundance of FDM, with only a small fraction of the total dark matter permitted to be in the form of FDM for the lowest particle masses, $m \leq \SI{e-25}{\eV} $ (e.\,g.\ \citealt{Hlozek2015, Rogers2023, Lague2024}).

While both CDM-only and FDM-only models face challenges, mixed dark matter (MDM) scenarios provide an alternative framework in which multiple dark matter species coexist and remain relatively unexplored. Here MDM refers specifically to mixed cold and fuzzy dark matter models. In particular, models combining CDM and FDM can retain CDM-like clustering on large scales while introducing partial smoothing on small scales \citep{Hu2000}. This reduces historically identified tensions without fully suppressing structure, and relaxes many of the constraints that apply to pure FDM models. Numerical simulations have demonstrated that the properties of collapsed structures depend strongly on the FDM fraction, including the strength of wave interference and the formation of central solitonic cores \citep{Schwabe2020}.

However, MDM does not necessarily resolve all small-scale issues. Low FDM fractions may not fully address problems such as core–cusp or too-big-to-fail, and there is a balance between achieving sufficient smoothing and avoiding the constraints associated with higher FDM fractions. In addition, the effects of FDM on the matter power spectrum can be degenerate with baryonic feedback \citep{Preston2025}. For our purposes, an additional complication arises in halo identification, as different dark matter components exhibit distinct physical behaviour. The presence of a wave-like component can smooth halo boundaries and complicate the definition of individual objects, requiring tailored halo-finding methods (see \cref{sec:HaloFinding}).

Here, we consider FDM fractions in the range $f = \numrange{0}{0.3}$, consistent with constraints from the Lyman-$\alpha$ forest (e.\,g.\ \citealt{Kobayashi2017, Irsic2017, Rogers2021, Wang2026}), high-redshift galaxy counts (e.\,g.\ \citealt{Ni2019}), reionisation history (e.\,g.\ \citealt{Bozek2015, Jones2021}), and CMB measurements (e.\,g.\ \citealt{Hlozek2018}). Exploring models in which CDM dominates but includes a subdominant FDM component provides a promising avenue for reconciling the predicted large- and small-scale behaviour with observations. Varying the FDM fraction also allows constraints to be placed on the viable parameter space of such models.

Recent work has constrained axions using the UV luminosity function from the Hubble Space Telescope \citep{Winch2024}. This model relies on assumptions about the shape of the halo mass function (HMF), so being able to strongly constrain axions in the mass range $ 10^{-26} \si{\eV}\leq m \leq 10^{-23} \si{\eV}$ will be vital for comparison with James Webb Space Telescope (JWST) observational data. In addition to modifying the abundance of haloes, FDM may also affect the internal baryonic evolution of collapsed structures through wave-driven dynamics that alter gas accumulation and fragmentation \citep{Tocher2026}. Accurate modelling of HMFs is therefore important not only for structure formation itself but also for connecting dark matter models to observable galaxy populations.

Simulations are required to study these scenarios because currently observational predictions for MDM cosmologies remain relatively underdeveloped, making it difficult to translate existing observations into constraints on the FDM fraction \citep{Ferreira2021}. Numerical simulations allow us to probe the physical effects of FDM and their environmental dependence (e.\,g.\ \citealt{Schwabe2016, Schive2014, Veltmaat2016}), and to track halo populations across redshift. However, the underlying physics of FDM is computationally demanding, requiring high-resolution grids and the solution of coupled Schrödinger–Poisson equations \citep[e.\,g.][]{Schwabe2016, 2021May}. Furthermore, the effects of small FDM fractions are subtle and difficult to detect observationally, particularly given current resolution limits. Even with future surveys, distinguishing these effects from other processes is likely to remain challenging.

Halo finders are therefore a crucial tool for analysing structure formation. A variety of methods exist (e.\,g.\ \texttt{Rockstar} \citep{Behroozi2013}, \texttt{SUBFIND} \citep{Springel2001}, \texttt{HBT+} \citep{Han2018}), but these typically assume that the matter distribution is particle-based and collisionless \citep{Knebe2011}. In MDM models, the presence of a grid-based, wave-like matter component complicates these assumptions, as halo boundaries become less well-defined and standard definitions may no longer apply.

Previous work has applied halo finders to MDM simulations by converting the FDM component into a CDM-like proxy \citep{Dome2025}, enabling the use of standard tools such as \texttt{Rockstar}. While this approach produces halo catalogues, it does not fully capture the wave nature of FDM or account for having different species of particles in the simulation. In contrast, here we reanalyse the simulations of \citet{Dome2025} using an FDM-aware methodology. We employ \texttt{AxiREPO}’s grid-outputting and grid-based halo finding to identify haloes directly from the combined density field.

Building on this, we develop a halo-finding pipeline for MDM simulations that consistently incorporates both CDM and FDM components. Using this framework, we construct halo catalogues and investigate the dependence of HMFs on FDM fraction and redshift. We then introduce a phenomenological transformation that maps CDM HMFs to their MDM counterparts. Our primary goal is to develop and validate a halo-finding framework for mixed cold–fuzzy dark matter simulations that can be applied more broadly. As applications, we study the evolution of HMFs with FDM fraction $f \leq 0.3$ and propose a mapping between CDM and MDM halo mass functions.

This paper is organised as follows. In \cref{sec:MDM} we discuss the underlying physics of our model. We discuss the simulation setup in \cref{sec:Simulations} and the halo finder in \cref{sec:HaloFinding}. The key findings on haloes and HMFs are discussed in \cref{sec:MDMHalos} and the transformation model from CDM to MDM is discussed in \cref{sec:TransformationFormula}. A general discussion of the significance and application of our results is found in \cref{sec:Discussion} before we conclude in \cref{sec:Conclusions}.

\section{Methodology}
In this Section we give some of the theoretical background, introducing MDM in \S~\ref{sec:MDM} and the simulations used in \S~\ref{sec:Simulations}, and describe halo finding in \S~\ref{sec:HaloFinding}.

\subsection{Mixed Dark Matter models}
\label{sec:MDM} 

The FDM fraction of a MDM model can be defined as
\begin{equation}
    f = \frac{\Omega_{\mathrm{FDM}}}{\Omega_{\mathrm{FDM}} + \Omega_{\mathrm{CDM}} + \Omega_{\mathrm{b}}}
\end{equation}
using the cosmological parameters from \cite{Planck2016}. This definition corresponds to the fractional contribution of FDM to the total matter density and is commonly adopted in studies of mixed dark matter cosmologies (e.\,g.\ \citealt{Lague2024, Dome2025}). Here, we focus on the dark matter-only results, so $\Omega_{\mathrm{b}} = 0$ for all simulations. Baryons are absorbed into the CDM component as they are assumed to behave similarly to CDM, such that $\Omega_{\mathrm{b}} \rightarrow 0 $ and $\Omega_{\mathrm{CDM}}$ is increased accordingly. Thus, in this case $f$ represents the ratio of FDM to total dark matter, such that the expression reduces to the more commonly used $f = \Omega_{\mathrm{FDM}}/(\Omega_{\mathrm{FDM}} + \Omega_{\mathrm{CDM}})$ (e.\,g.\ \citealt{Vogt2023}). This means that $f = 1$ corresponds to an FDM-only model.

In the simulations, FDM is considered to be analogous to an axion or axion-like particle. The axion arises from the Peccei-Quinn mechanism, originally proposed as a solution to the strong CP problem in quantum chromodynamics \citep{PecceiQuinn1977a,PecceiQuinn1977b}, with the associated particle subsequently identified by \citet{Weinberg1978} and \citet{Wilczek1978}. Axions were subsequently recognised as a potential dark matter candidate, produced through coherent oscillations of the axion field in the early Universe (see e.g. \citealt{Preskill1983}, \citealt{ChadhaDay2022}). FDM particles are a popular dark matter candidate because they are stable, weakly interacting and long-lived particles \citep{Sikivie2024}. The particles can be parametrised by an angular variable, $\theta$, which represents the misalignment of the axion-like particle field $\phi$, with the field decay constant $f_{\mathrm{a}}$:
\begin{equation}
    \theta = \frac{\phi}{f_{\mathrm{a}}} .
\end{equation}
The initial random value of this misalignment angle determines the FDM abundance in the simulation.

$\phi$ is periodic and so we can construct a complex scalar field to represent the classical wavefunction (in natural units):
\begin{equation}
    \psi = \sqrt{\rho_{\mathrm{FDM}}}e^{i\gamma} , 
\end{equation}
where $\rho_{FDM}$ is the FDM density and $\gamma$ is the phase of the complex scalar field.
The Schrödinger equation (in natural units) and Poisson equations then apply:
\begin{equation}
    i\mleft(\partial_{t} + \frac{3\dot{a}}{2a}\mright)\psi = \mleft(-\frac{1}{2m}\nabla^2 + m\Phi\mright) \psi ,
    \label{SP1}
\end{equation}
\begin{equation}
    \nabla^2\Phi = 4\pi G(\rho_{\mathrm{CDM}} + \rho_{\mathrm{FDM}} - \bar{\rho}_{\mathrm{tot}}) ,
    \label{SP2}
\end{equation}
where $\Phi$ is the Newtonian gravitational potential, $\rho_{\mathrm{FDM}} = |\psi|^2$, and $\bar{\rho}_{\mathrm{tot}}$ is the mean total dark matter density. 

Defining the comoving quantities
\begin{equation}
    \psi_{\rm c} = a^{3/2}\psi, \qquad
    \rho_{i,\rm c} = a^3 \rho_i, \qquad
    \nabla_{\rm c} = a\nabla, \qquad
    \Phi_{\rm c} = a\Phi,
\end{equation}
the Schrödinger--Poisson equations become
\begin{equation}
    i\frac{\partial \psi_{\rm c}}{\partial t}
    =
    -\frac{1}{2ma^2}\nabla_{\rm c}^2\psi_{\rm c}
    + \frac{m}{a}\Phi_{\rm c}\psi_{\rm c},
\end{equation}
and
\begin{equation}
    \nabla_{\rm c}^2\Phi_{\rm c}
    =
    4\pi G
    \left(
        \rho_{\rm CDM,c}
        + \rho_{\rm FDM,c}
        - \bar{\rho}_{\rm m,c}
    \right).
\end{equation}
These are the comoving Schrödinger--Poisson equations evolved by
\texttt{AxiREPO}.

This system describes the non-linear evolution of a self-gravitating scalar field in the classical, non-relativistic limit within an expanding Universe, with the gravitational potential sourced by both the FDM and CDM density fields.

\subsection{Simulation Setup}
\label{sec:Simulations} 

The simulations analysed in this work were originally presented by \citet{Dome2025} and were performed using \texttt{AxiREPO} \citep{2021May,May2023}. \texttt{AxiREPO} evolves the FDM component as a wavefunction satisfying the Schr\"odinger-Poisson equations, while the CDM component is represented by collisionless $N$-body particles. The two components are coupled gravitationally through the total matter density and therefore evolve self-consistently within the same gravitational potential. The FDM wavefunction is evolved on a uniform Cartesian mesh using a second-order symmetrised split-step pseudo-spectral Fourier method, while the CDM component is evolved using the particle-based gravity solver.

The initial conditions were generated at $z=127$ using \texttt{MUSIC} \citep{Hahn2011}. Separate density and velocity transfer functions for the CDM and FDM components were calculated using a modified version of \texttt{AxionCAMB} \citep{Hlozek2015}. These component-specific transfer functions account for the scale-dependent suppression and growth of the FDM density field. The resulting FDM density and velocity fields were used to construct the initial complex wavefunction through the Madelung representation. Further details of the initial-condition generation and numerical implementation are given by \citet{Dome2025}.

These simulations were performed in 60 cMpc/h boxes using an underlying $2048^{3}$ grid on which the $m = 10^{-24.5}\,\si{\eV} = \SI{3.16e-25}{\eV}$ FDM component was evolved. The FDM fractions tested were \numlist{0.01; 0.1; 0.2; 0.3}. The CDM component was represented by $N$-body particles, as is standard in cosmological simulations. The redshift range explored was between $ z \simeq 1$ and $z \simeq 10$, but for low FDM fractions ($f\leq0.1$) the calibration of the MDM model with \texttt{AxionHMcode} \citep{Vogt2023} is only reproducible over the range $1 \leq z \leq3.5$. For redshifts below $z = 1$ the highest velocity dispersions may remain unresolved by the FDM solver, hence these redshifts are generally excluded from analysis. The MDM simulation was evolved to $z = 0$, but the validity of the FDM evolution in this regime is less well-established, so results from $0 \leq z \leq1$ are not included in our investigation.

Due to the dominant CDM component, the potential wells become steeper than in a \SI{100}{\percent} FDM case, which increases the FDM velocity dispersion, thus decreasing its de Broglie wavelength. The MDM simulations thus require higher resolution than pure FDM counterparts.

We also require a consistent and physically motivated definition of a halo. In MDM, we define haloes as connected regions of the total dark matter density field that exceed a specified overdensity threshold. This choice is motivated both by the numerical representation of the MDM system as a density field and by the underlying physical definition of a halo as a gravitationally bound overdensity. Unlike some other halo finders, we do not require virial equilibrium explicitly in our haloes and do not check for comoving components in the halo e.\,g.\ in the style of a 6D halo finder. This is because in FDM there is no 6D phase space.\footnote{Although analogues like the Husimi or Wigner distributions can be defined as in quantum mechanics, these are only quasi-phase space distributions and not directly equivalent to the particle case.}

\subsection{Halo Finder}
\label{sec:HaloFinding} 

\subsubsection{Grid Construction}
\label{sec:GridConstruction} 
\begin{figure*}
    \centering
	\includegraphics[width=1.0\textwidth]{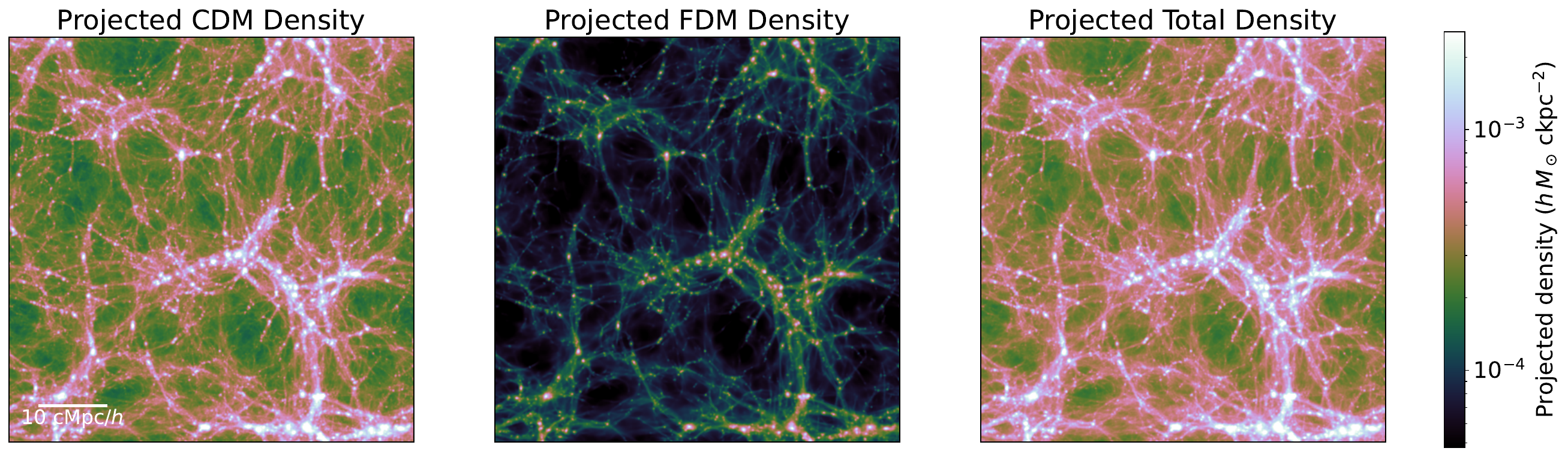}
    \caption{(Comoving) density grids for the CDM component (left), FDM component (middle), and total density (right) for the $f = 0.2$ model at $z=2$. The CDM and FDM distributions can be seen to be very similar with the density nodes appearing at the same locations in both.}
    \label{fig:density grids}
\end{figure*}

As FDM uses grid-based numerical methods, the FDM component is represented as a wavefunction defined on a grid. However, the CDM component is expressed as discrete particles in unconstrained locations. This combination of particle and grid components makes it difficult to directly sum the overdensities in a multi-species MDM model. To counteract this we first modified the CDM component to be in the form of a grid of the same size and dimensions as the FDM component to allow them to be directly combined (\cref{fig:density grids}).

\subsubsection{Grid-based halo finder}
In \texttt{AREPO}, haloes are traditionally identified using a friends-of-friends (FOF) method, which connects simulation particles to other particles within a specific threshold distance. However, in our entirely grid-based representation, there are no particles and haloes must be detected by looking for overdensities on the Cartesian mesh. The grid-based halo finder implemented in \texttt{AxiREPO} utilises a modified version of the FoF algorithm for a discretised density mesh \citep{2021May, May2023}. This halo finder uses a density threshold as a parameter rather than the traditional linking length. The density threshold corresponding to a halo is a user-specified parameter. If the density of a cell exceeds the threshold, it is then linked to adjacent cells. This is all self-contained within the \texttt{AxiREPO} FDM code. This halo finder has been extensively validated in CDM-only simulations as described in \citet{2021May}, where it was shown to reproduce standard FOF halo catalogues with good accuracy (their Figure~5). 

For each halo identified by the grid-based finder, we characterise its mass using \(M_{200c}\), where \(R_{200c}\) is defined as the radius within which the mean enclosed density is 200 times the critical density of the Universe at that redshift, and \(M_{200c}\) is the total enclosed mass within this radius.

Since our implementation involves assigning CDM particles onto a grid and combining this with the FDM density field, it is important to verify that the halo identification remains consistent in this mixed-species context. We find that the grid-based halo finder reproduces the CDM halo mass function obtained using the FOF method at a level of 10 per cent or better over the well-resolved mass range (\cref{fig:rockstaroverpredict}). While no direct comparison is possible in the MDM case due to the absence of a standard multi-species halo finder that treats FDM self-consistently, the agreement in the CDM limit, combined with the stability of the HMF under variations of the overdensity threshold (\cref{fig:halofinderthreshold}), yields stable relative differences between CDM and MDM cosmologies, which is the main focus of this work.

\subsubsection{Setting the overdensity criterion}
One of the user-determined parameters is the value of the overdensity criterion used to define a halo. This represents how much denser a region needs to be than the average background density to be counted as a halo.

In the context of the HMF, the overdensity criterion represents the threshold between the more diffuse cosmic web structure formed by FDM and the discrete, more globular dark matter halo. As the FDM traces the CDM, and the simulations are majority CDM, the largest haloes appear in similar locations to those in a CDM-only run, but the properties of individual haloes can vary slightly.

\begin{figure}
	\includegraphics[width=\columnwidth]{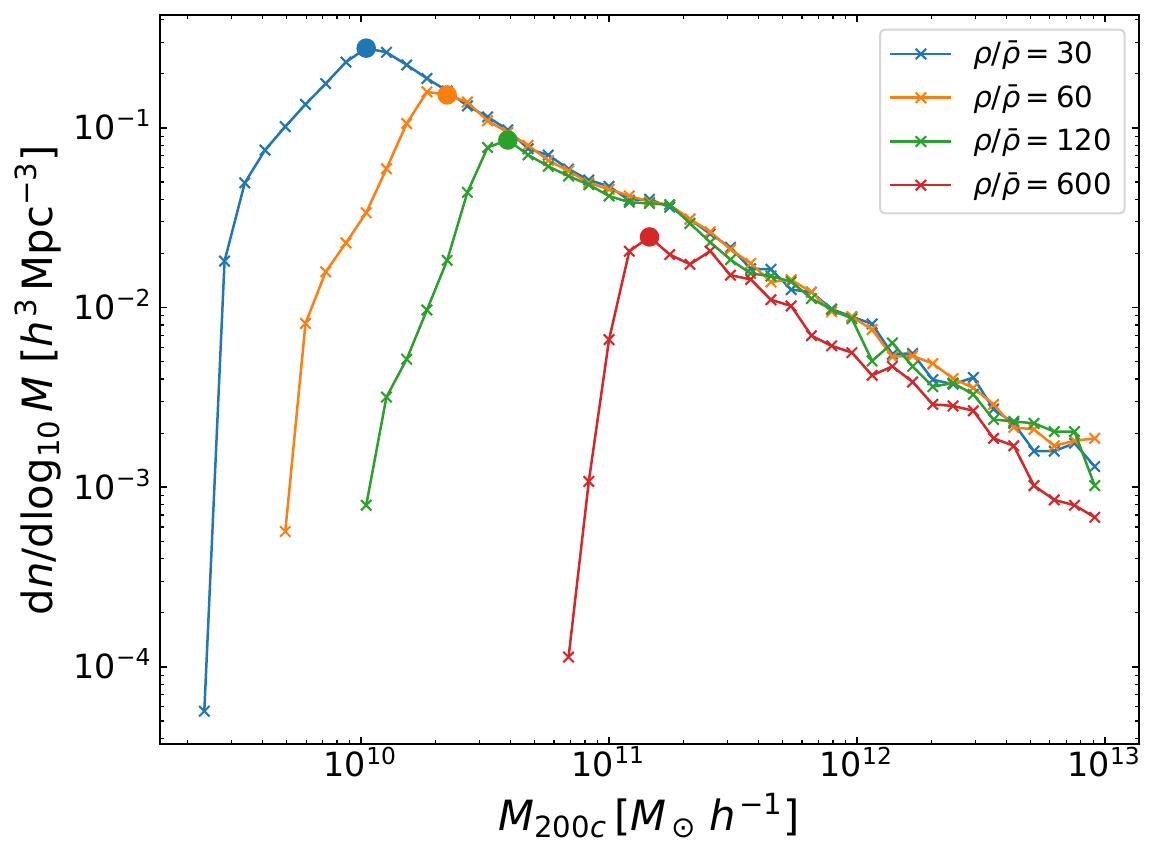}
    \caption{HMFs using the $M_{200c}$ mass definition for different overdensity thresholds for the grid-based halo finder, shown for the $f=0.1$ case at $z=1$. Filled circles mark the resolution turnover for each threshold. The shape of the HMFs is broadly similar, but the turnover shifts to higher masses for higher thresholds. This demonstrates that the overdensity threshold affects the absolute turnover mass. As shown by the red curve, using a very high value of the threshold limits the number of objects detected across \textit{all} masses. At this very high threshold, many low-mass haloes are not identified, while the catalogue also becomes increasingly sensitive to how connected high-density regions are partitioned into individual objects. The fiducial value adopted throughout this work is $\rho/\bar{\rho}=60$.}
    \label{fig:halofinderthreshold}
\end{figure}

The impact of varying the overdensity criterion can be seen in \cref{fig:halofinderthreshold}. The following behaviour reflects the numerical properties of the grid-based halo finder rather than physical changes in the underlying cosmology. In particular, the low-mass resolution turnover in the HMF arises from the combination of the finite grid resolution and the adopted overdensity threshold. Similar sensitivities to the halo-definition criterion are also present in traditional halo finders, for example through variations in the FOF linking length. For sufficiently low overdensity thresholds, the recovered HMF therefore approaches a similar form to that obtained from a traditional FOF halo finder. Increasing the overdensity criterion moves the HMF turnover to higher mass bins. There are some small differences in the lowest-mass haloes as these become resolution-limited in the grid-based method because their density is always spread out over at least the volume of one grid cell. There are similar statistical anomalies for the highest-mass haloes as here there are only a few haloes per bin, so finder-specific choices about whether large haloes count as multiple massive objects or one very massive object lead to more noticeable changes in the HMF. 

We quantify the sensitivity of the HMF to the adopted overdensity threshold by measuring the shift in the low-mass turnover for different values of $\rho/\bar{\rho}$. Varying the threshold from $\rho/\bar{\rho}=30$ to $120$ shifts the turnover mass by $0.57\,\mathrm{dex}$, demonstrating that the absolute HMF, and in particular its low-mass completeness limit, depends on the adopted halo-finder criterion. This behaviour reflects the numerical definition of haloes in the grid-based finder rather than a change in the underlying cosmology.

All results in this work use a fixed fiducial overdensity threshold of $\rho/\bar{\rho}=60$, ensuring that the same halo definition is applied consistently to the CDM and MDM simulations. As the primary quantity used later in this work is the relative MDM-to-CDM HMF suppression, some of the halo-definition dependence is expected to cancel when the two cosmologies are compared using the same criterion. We quantify the residual sensitivity of the suppression ratio to the overdensity criterion explicitly in Section~4.2.

\subsubsection{Halo finding between different species}

\begin{figure}
	\includegraphics[trim={0.2cm 0.2cm 0.2cm 0.2cm}, clip, width=\columnwidth]{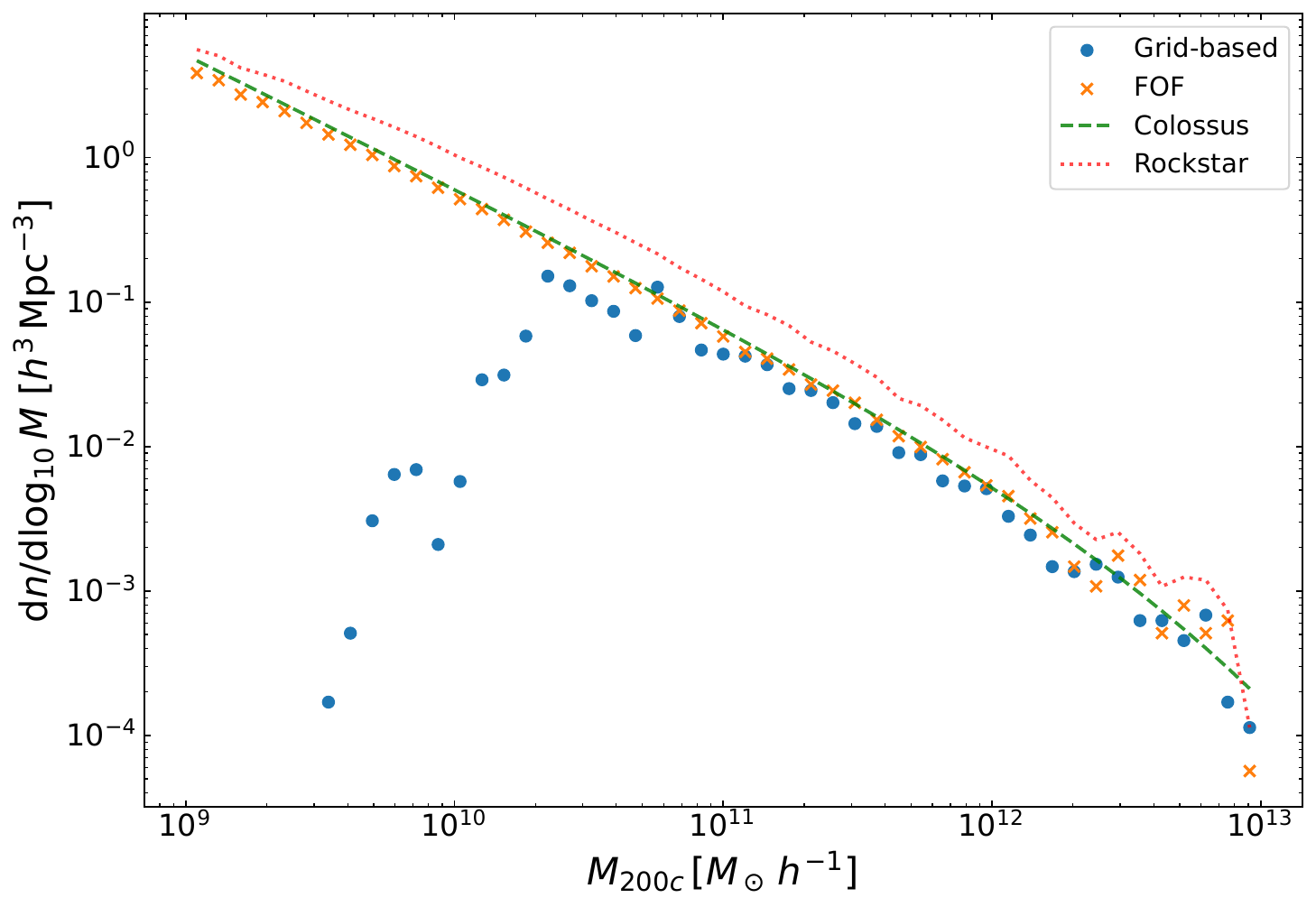}
    \caption{HMF for the CDM-only case using $M_{200\mathrm{c}}$ for the \texttt{AREPO} FOF halo finder (orange crosses), the \texttt{AxiREPO} grid-based halo finder (blue circles), the \texttt{Colossus} \citep{Diemer2018} Tinker analytical model \citep{Tinker2008} (green dashes) and \texttt{Rockstar} halo finder (red dotted line) for $z=2$. The FOF and grid-based halo finders are consistent. The CDM HMF agrees with the \texttt{Colossus} Tinker model at the $\sim \SI{10}{\percent}$ level across the fitted mass range. The \texttt{Rockstar} results lie systematically above the other catalogues.}
    \label{fig:rockstaroverpredict}
\end{figure}

In previous studies (e.\,g.\ \citealt{Anderhalden2012}) halo finding in MDM simulations was performed either i) independently on each species of dark matter or ii) by approximating one species of dark matter to be like the other. In \cite{Dome2025}, FDM was approximated as CDM proxy particles which allowed traditional one-species halo finders like \texttt{Rockstar} to be used.
While this choice allowed halo finders to operate on MDM scenarios, it has always come at the cost of either not detecting joint haloes or making assumptions about the behaviour of one of the species of dark matter, e.\,g.\ ignoring the wave-like behaviour of FDM when treating it as CDM.

In \cref{fig:rockstaroverpredict}, we show that in the \cite{Dome2025} implementation the \texttt{Rockstar} HMF lies systematically above the other HMFs compared to analytical models and other halo finders. This figure shows the \texttt{AREPO} traditional FOF halo finder (orange crosses) and the grid-based halo finder (blue circles) that we use. It can be seen that the FOF and grid-based halo finders show good agreement, until the resolution turnover for the grid-based finder. Both of these halo finders also show good agreement for this CDM case with the \texttt{Colossus} model (green dashes), while the \texttt{Rockstar} model (red dotted line) is shown to lie above the other curves across all mass ranges. This is shown to be true in the CDM-only case and \texttt{Rockstar} alone does not have the capability to do halo finding on FDM. The treatment of the smooth wave-like FDM component as discrete particles may alter halo identification and substructure assignment, potentially biasing the inferred level of small-scale suppression. This effect could become even more prominent in MDM cases.

Using our joint grid-based method (see \cref{sec:GridConstruction}), we can directly find haloes as an overdensity in the total amount of dark matter, rather than local overdensities in each dark matter species. This combined treatment of the haloes also allows us to investigate what happens at the boundaries of MDM haloes as the presence of FDM can lead to some smoothing of small-scale structures.

Overall, applying a single halo-finding method that treats both components consistently provides a coherent framework for analysing MDM simulations.

\section{Results: Mixed Dark Matter haloes}
\label{sec:MDMHalos} 

This section presents the first part of our results. We start in \S~\ref{sec:FDMtracing} by showing how FDM traces the CDM component, then continue in \S~\ref{sec:FDM impact} by comparing the appearance of the cosmic web in CDM and FDM simulations, as a function of time and the fraction of FDM. Then we make a more quantitative comparison by examining the impact on the form of the HMF in \S~\ref{sec:HMFs}.

\subsection{The large-scale correlation between CDM and FDM density fields}
\label{sec:FDMtracing}

\begin{figure}
	\centering
	\includegraphics[width=\columnwidth]{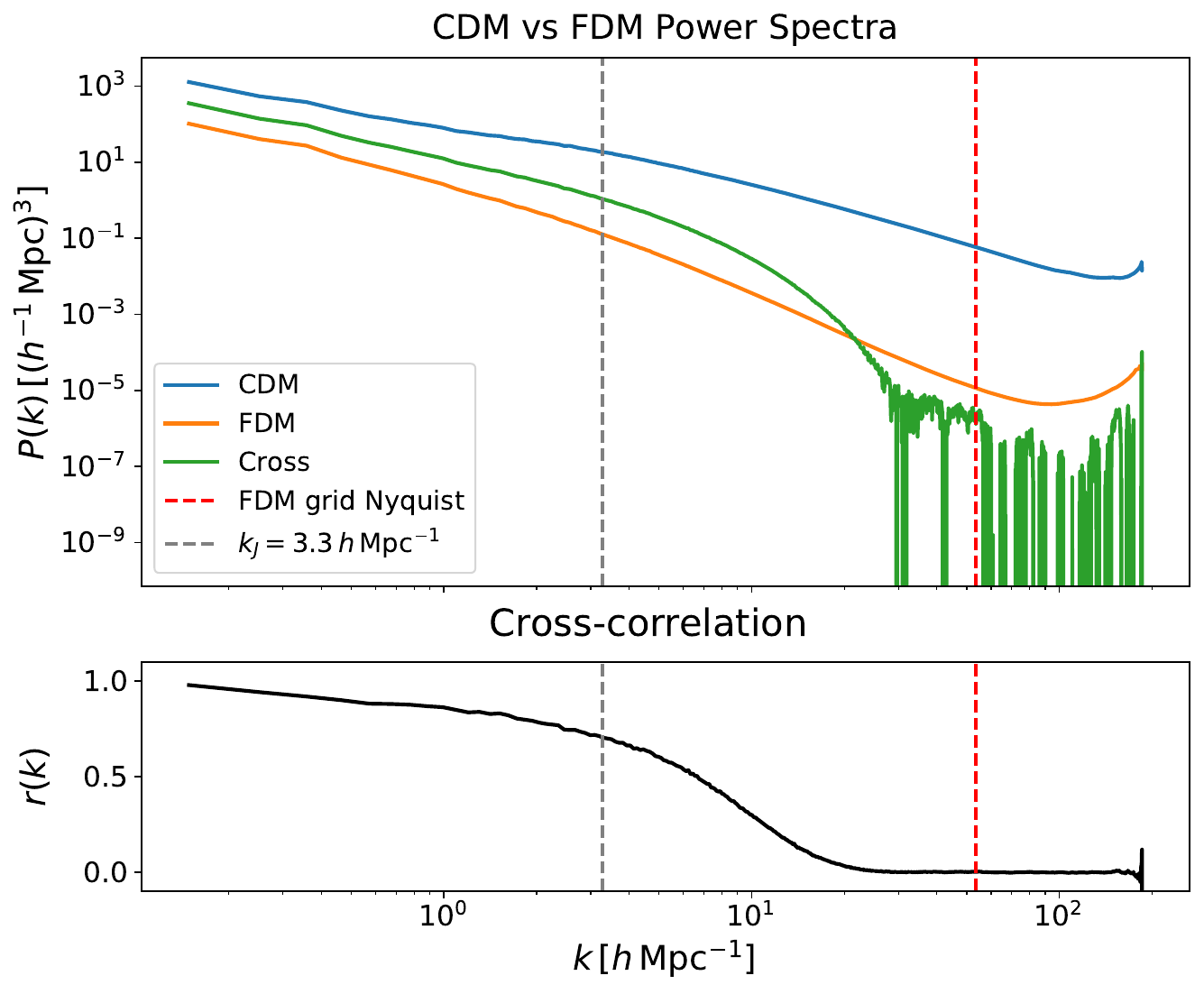}
    \caption{The top plot shows auto power spectra for CDM (blue) and FDM (orange) and the cross power spectrum of CDM and FDM (green) for the $f=0.1$ cosmology at $z=1$. The lower plot shows the cross-correlation of the CDM and FDM, where $r(k) = 1$ is perfect correlation. The red line shows the FDM grid Nyquist wavenumber, hence results to the right of the red line are dominated by resolution issues. The grey line shows the theoretical FDM Jeans scale.}
    \label{fig:crosspowerspec}
\end{figure}

We begin by quantifying the extent to which the CDM and FDM density fields trace each other. In \cref{fig:crosspowerspec}, we present the auto-power spectra of the CDM (blue) and FDM (orange) fields, as well as their cross-power spectra (green). The lower panel shows the cross-correlation coefficient, $r(k)$, which shows how similar the fields are at each value of $k$ by comparing the Fourier modes of CDM and FDM and measuring whether they have the same phase and structure. The cross correlation coefficient is described by
\begin{equation}
    r(k) = \frac{P_{\mathrm{cross}}(k)}{\sqrt{P_{\mathrm{CDM}}(k)P_{\mathrm{FDM}}(k)}} .
    \label{rk}
\end{equation}

The cross power spectrum of the CDM and FDM fields shows strong correlation between the CDM and FDM on large scales, showing that the FDM traces the CDM at these scales. This is consistent with the similarity of the large-scale structure between the CDM and FDM fields. At small scales (high $k$), the cross power spectrum is very noisy with large fluctuations. This could be due to physical differences between the CDM and FDM models which apply at large $k$, where FDM loses small-scale structure. However, this could also be due to the resolution limit and numerical noise in our simulations or due to low absolute power spectrum levels. At small scales, we would expect CDM to still have granular structure, whereas FDM smooths out the smallest structure, so the fields no longer match. The spike at the highest $k$ values is attributed to numerical noise due to division by a very small number.

The linear FDM transfer function used to construct the initial conditions contains the expected oscillatory behaviour at high \(k\). However, these oscillations occur on scales that are already strongly non-linear by \(z=1\). Non-linear growth and mode coupling generate substantial small-scale power, overwhelming the strongly suppressed linear FDM contribution and obscuring the detailed oscillatory structure of the initial transfer function. At the highest wavenumbers, proximity to the FDM-grid Nyquist scale further limits the reliability of the measured power spectrum.

There is an intermediate region of the plot where, as $k$ increases, the cross power spectrum drops faster than the auto power spectra which indicates a transition region where the FDM field no longer matches the CDM distribution as well and the two fields become decorrelated. This transition occurs around scales where FDM wave effects become important and is related to the particle-mass-dependent Jeans scale.

These results show that FDM traces CDM on large scales and forms the same cosmic web structure. This is because at these scales gravity dominates and so wave effects are negligible.

\subsection{FDM impact on haloes}
\label{sec:FDM impact} 

\begin{figure*}
	\centering
	\includegraphics[trim={2cm 3cm 2cm 2cm}, clip, width=\textwidth]{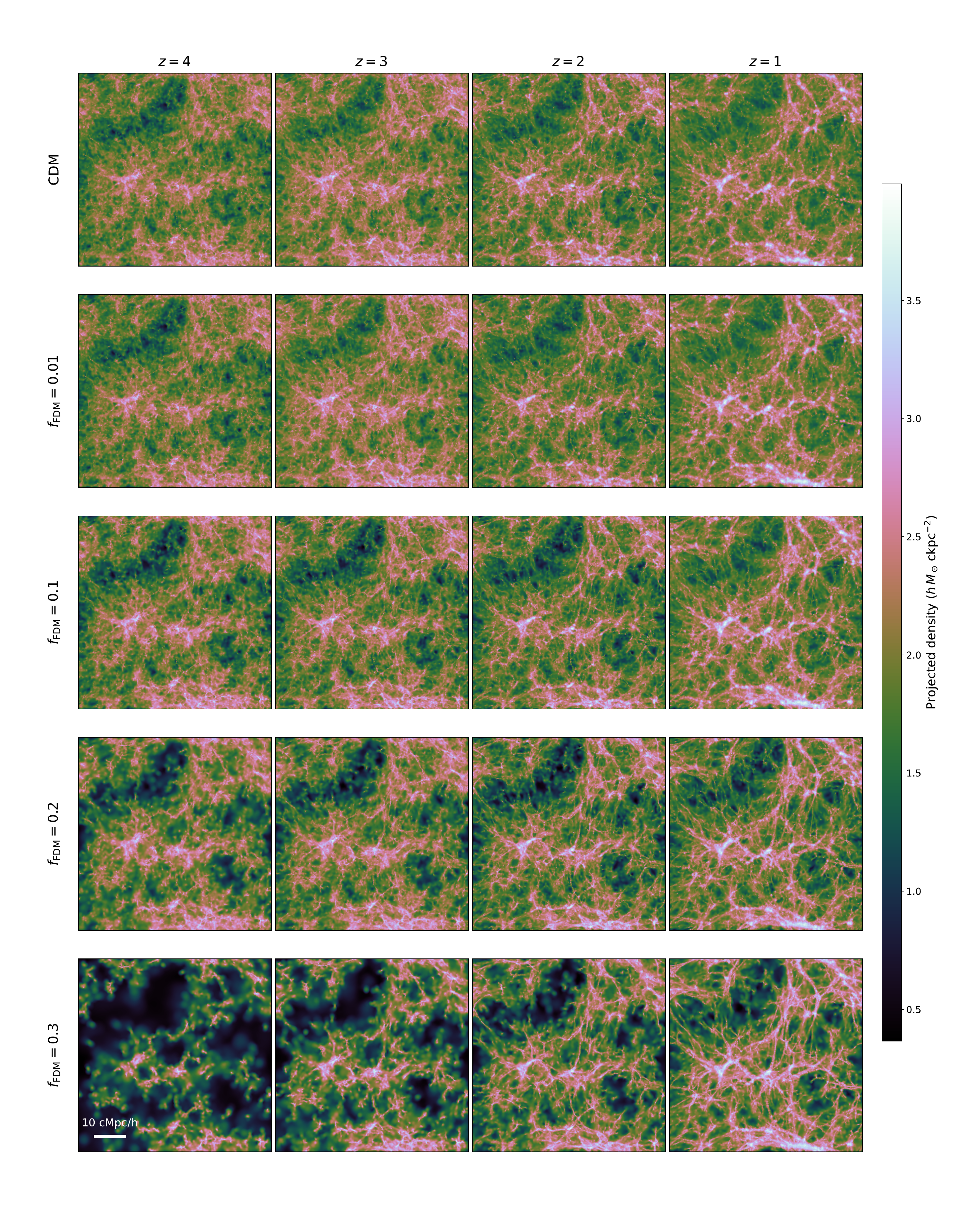}
    \caption{Large-scale structure within the simulation box, shown for four redshifts (columns from left to right $z = 4, 3, 2, 1$) and five cosmologies (rows from top to bottom, CDM, $f= 0.01, f=0.1, f = 0.2, f =0.3$). The differences in the granularity and formation of structures can be seen with increasing FDM fraction (downwards) and the effect on the resulting structures over time can be seen with decreasing redshift (to the right). The colourbar is normalised in the same way across all of these figures.}
    \label{fig:cosmicwebs}
\end{figure*}

A comparison of cosmic web structures for different FDM fractions and redshifts can be seen in \cref{fig:cosmicwebs}. Each of the rows shows one cosmology (CDM, $f=0.01$, $f=0.1$, $f=0.2$, $f=0.3$, from top to bottom) and the columns show the redshift evolution from $z = 4$ to $z = 1$ (from left to right). The colourbar is the same across the figure allowing for direct visual comparison between the models. Increasing the fraction of FDM inhibits the formation of small-scale structures, leading to a more diffuse and smoother filamentary structure. Small-scale, granular structures are less prominent in the MDM cases due to the structure smoothing. The voids also appear more prominent for a given redshift when comparing CDM and MDM due to this delayed structure formation. Voids have been shown to be a useful probe of the nature of dark matter and can be used to place constraints on axion dark matter models \citep{Dome2023dissection, London2026}.

\begin{figure*}
	\centering
	\includegraphics[width=\textwidth]{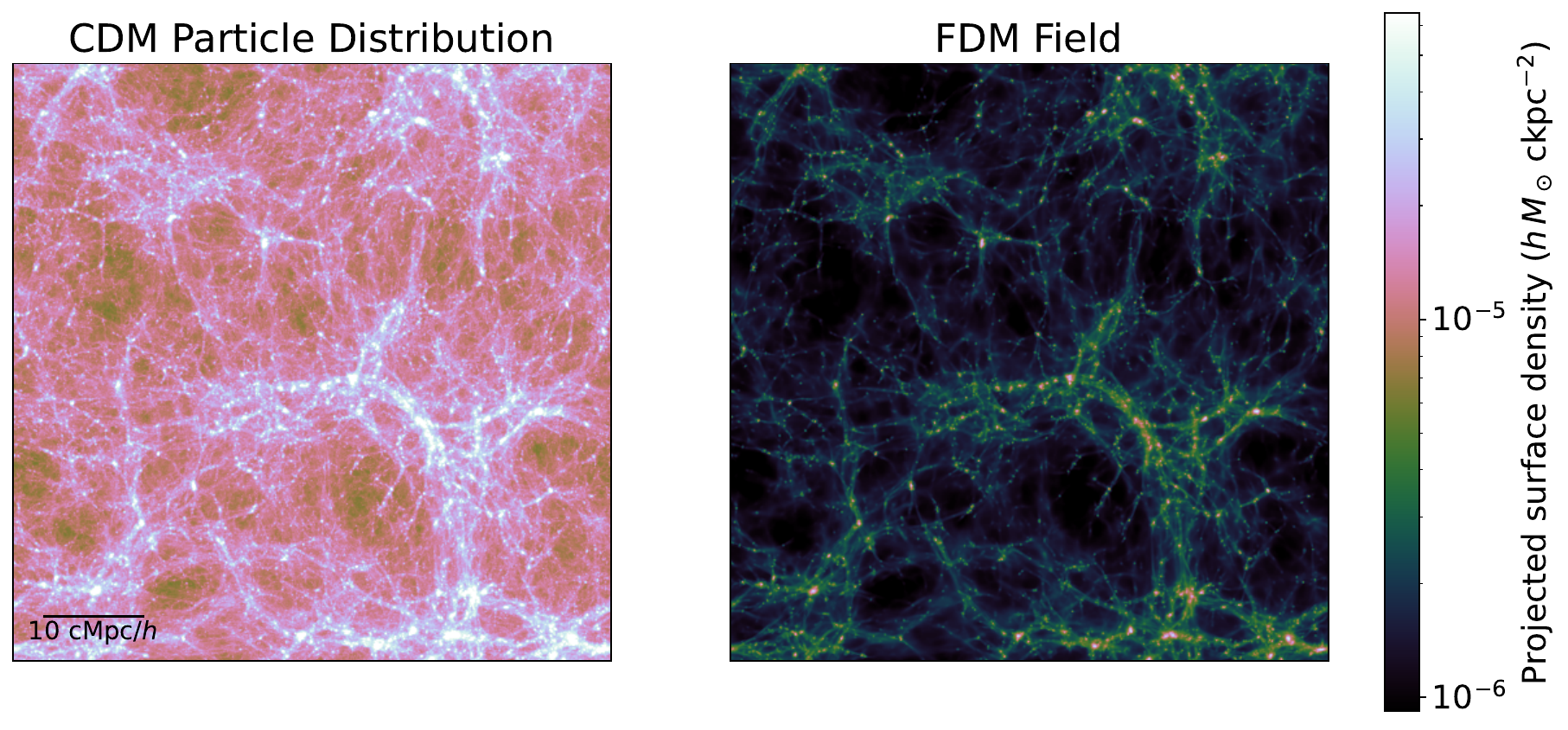}
    \caption{CDM particle distribution (left) and FDM field (right) for the $f=0.1$ cosmology at $z=2$. The highest-density parts of the FDM field can be seen to be at the same locations as the most dense areas where many CDM particles are found. This shows the FDM distribution traces the CDM. This visual agreement is consistent with the quantitative cross-correlation analysis shown in \cref{fig:crosspowerspec}.}
    \label{fig:cdmfdmtrace}
\end{figure*}

However, overall the structure produced looks very similar across all the cases with the largest haloes forming in the same locations. This arises because the FDM traces the CDM component (\cref{fig:cdmfdmtrace}) so where the CDM overdensities are will also be where the FDM gathers. As all MDM models are CDM-dominated, the large-scale CDM distributions remain very similar and the largest haloes tend to form in corresponding locations.

FDM is expected to smooth small-scale density structure physically through wave effects. However, the grid-based halo finder may additionally enhance this behaviour because smoother density distributions produce less sharply defined halo boundaries. This is a notable issue within alternative dark matter models across the field, and often specific conditions have to be applied to ensure distinctions for the edges of haloes and to categorise cosmic web features e.\,g.\ where a void becomes a filament \citep{May2023}.

\begin{figure*}
	\centering
	\includegraphics[trim={2cm 3cm 2cm 2cm}, clip, width=\textwidth]{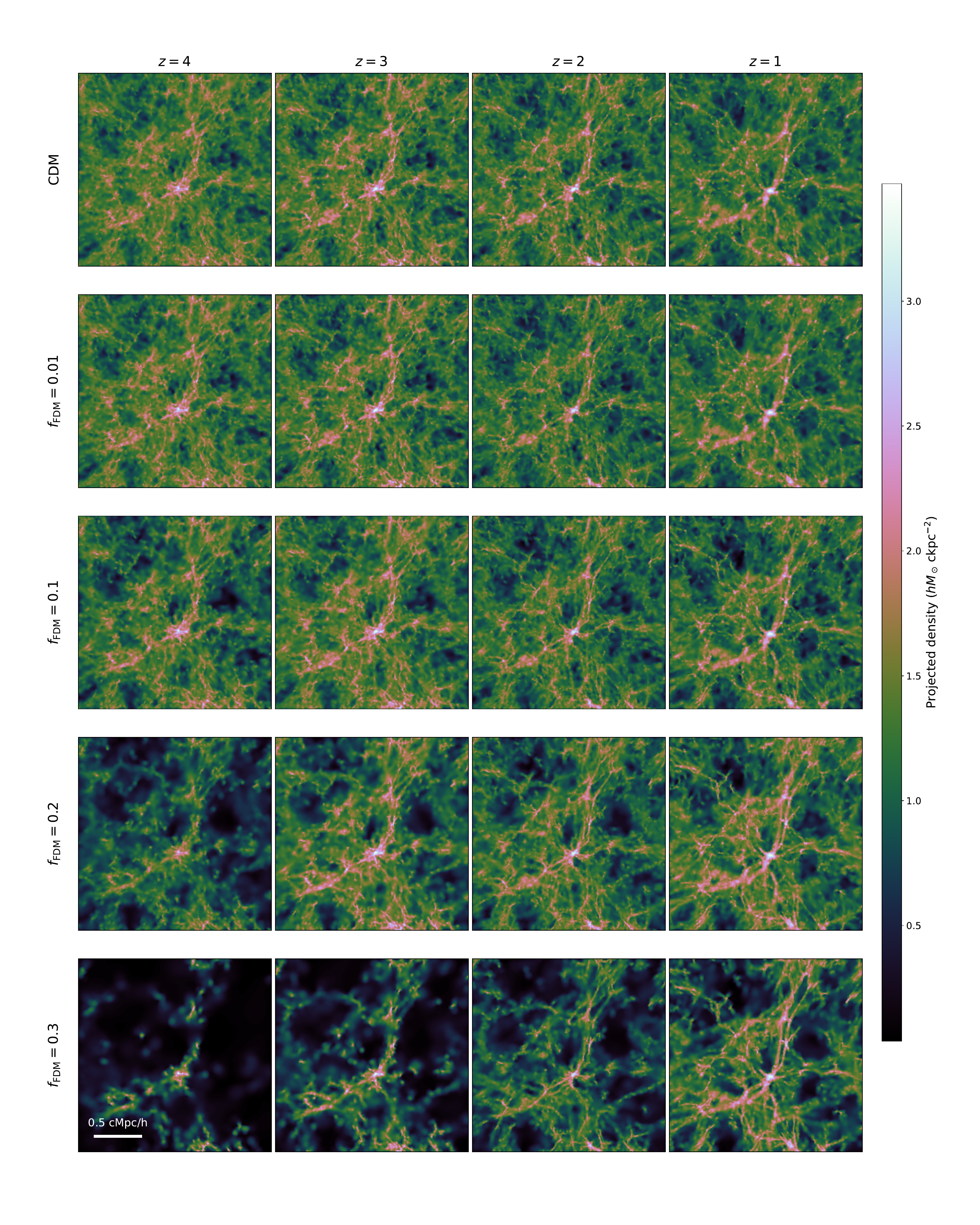}
    \caption{A zoom-in of one of the largest haloes for redshifts ($z = \numlist{4; 3; 2; 1}$) and five cosmologies (CDM, $f = \numlist{0.01; 0.1; 0.2; 0.3}$). There can be seen to be reduced small-scale, granular structure in the highest FDM fraction compared to the lower fractions which are in environments abundant with smaller structures.}
    \label{fig:largesthalo}
\end{figure*}

Given our focus on statistical differences between cosmologies, we validate the halo catalogues through the HMF rather than through one-to-one halo matching, which provides a robust measure of the abundance of collapsed structures independent of individual halo matching. This approach is commonly adopted when comparing halo definitions across different finders or physical models. We can compare the CDM HMFs to our MDM models to see the distribution of haloes in the model. We can also look at our highest-mass objects, i.\,e.\ our largest haloes for comparison across the simulations (\cref{fig:largesthalo}).

\subsection{The evolution of the halo mass function}
\label{sec:HMFs} 

\begin{figure*}
	\centering
	\includegraphics[width=\textwidth]{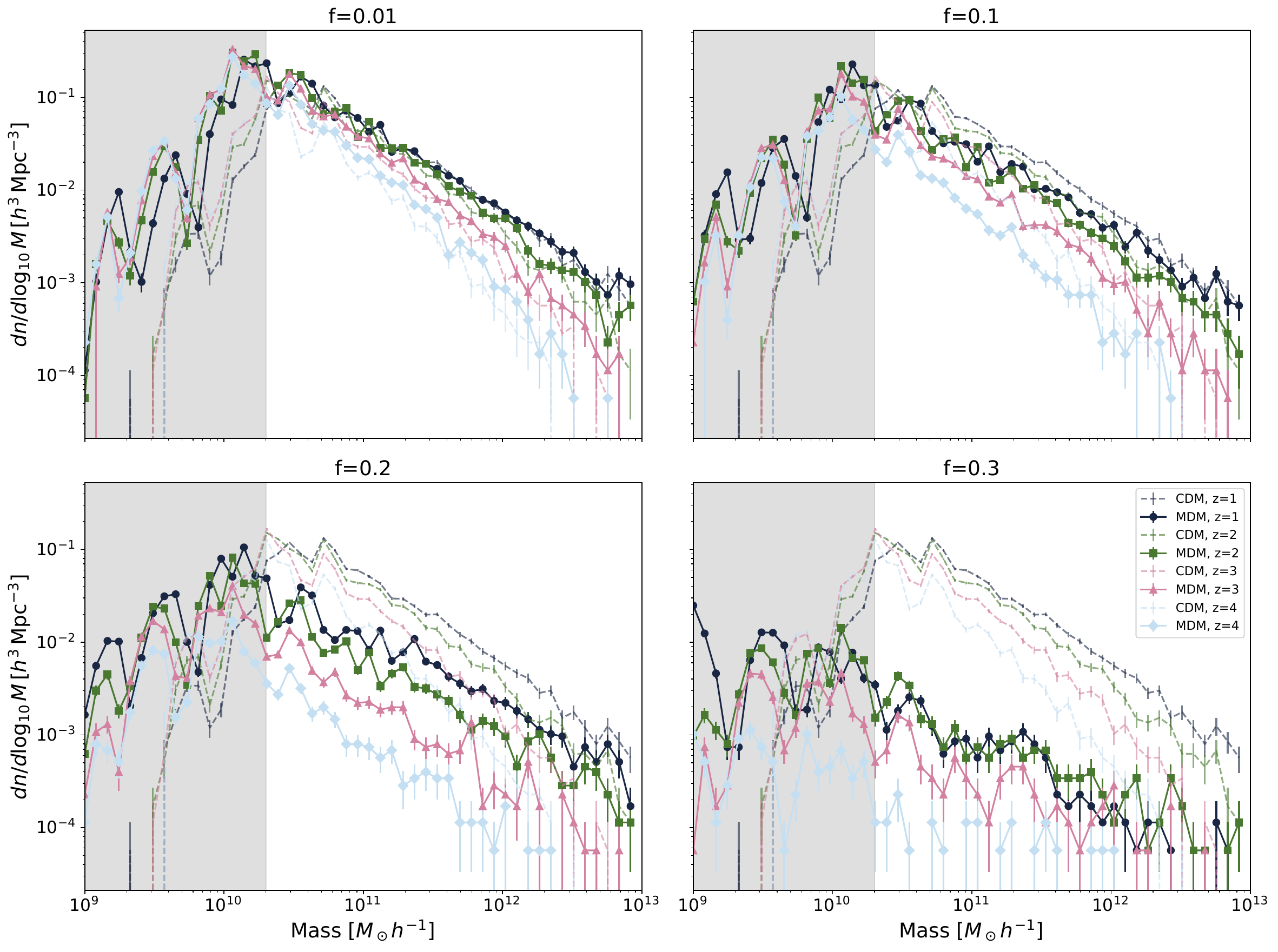}
    \caption{Halo mass functions using $M_{200\mathrm{c}}$ shown for each fraction of FDM (as given by the panel title) across the redshift range $z = \numrange{1}{4}$. The grey shaded region is the region excluded from the fitting analysis because it lies below the adopted resolution cut. Error bars show HMF uncertainties propagated from Poisson halo-count uncertainties. Dashed lines show the \SI{100}{\percent} CDM cosmology results at the same redshifts. The CDM curves are identical across panels. The MDM values are lower across all bins with increasing FDM fraction as there are fewer haloes due to the delayed structure formation. The slope of the HMF can also be seen to be shallower. The redshift evolution broadly follows the hierarchical growth expected in CDM, with the abundance of high-mass haloes increasing towards lower redshift.}
    \label{fig:HMFswithz}
\end{figure*}

As shown in \cref{fig:HMFswithz}, the measured HMF evolves systematically with both redshift and FDM fraction. Increasing the FDM fraction progressively suppresses the low-mass end of the HMF while also reducing the overall halo abundance. This behaviour is consistent with delayed structure formation caused by FDM wave effects, which suppress the growth of small-scale density fluctuations and therefore delays the formation and growth of low-mass haloes. As a result, increasing the FDM fraction produces progressive suppression and flattening in the HMF.

At higher FDM fractions, we additionally observe a flattening of the high-mass end of the HMF, producing a feature that we refer to as a \enquote{high-mass flattening}. The origin of this behaviour is not fully clear and may reflect both physical effects and properties of the halo-finding procedure. Since haloes in the grid-based method are identified as connected overdense regions, nearby massive structures may occasionally be merged into a single larger object. As an illustrative comparison, we examined the positions of the largest haloes identified using the grid-based and FOF methods and found cases where a single high-mass object in the grid-based catalogue was resolved as multiple haloes in the FOF catalogue. This comparison is intended only as a qualitative example and does not constitute a systematic analysis of the origin of the high-mass behaviour. 

All HMFs show growth towards lower redshift in the number of high-mass haloes as lower-mass haloes evolve over time and grow in mass. This arises because, regardless of the FDM fraction, we would expect that once the critical density for a halo to form is reached that halo would continue to accrete mass. As structures continue to merge and accrete material over time, the abundance of high-mass haloes increases towards lower redshift.

\section{A fitting formula for the halo mass function}
\label{sec:TransformationFormula}

\subsection{Motivation}
While having halo-finding methods for specific MDM simulations is important for improving our understanding of the structures that form in these cosmologies, a broader consequence of our work is that it provides a transformation between CDM and MDM HMFs as a function of halo mass, FDM fraction, and redshift. This provides a useful framework for investigating how changes in the underlying matter distribution are reflected in the abundance of dark matter haloes. Since the HMF is sensitive to the shape and amplitude of the linear matter power spectrum, measurements of halo abundances may be used to constrain models of small-scale power suppression and the allowed FDM fraction. However, the mapping between the linear power spectrum and the HMF is nonlinear and non-unique, so such constraints require forward modelling rather than a direct inversion of the HMF.

In our particular case, we have explored a set of discrete snapshots from MDM models, i.\,e.\ simulation snapshots are available only at discrete combinations of FDM fraction and redshift. While this already provides information about a region of the parameter space, having a model for the HMF for any given FDM fraction and redshift allows for more extensive investigation. 

In \cite{Dome2025}, the authors determine a two-parameter model of the form of \cite{Schive2016}  where they fix one parameter, $\alpha(f,z)$, adopted directly from a pure FDM model, and tune another parameter, $\beta(f,z)$. Here we extend this to a more flexible model which encodes the suppression mass scale and transition shape as a linear function of $f$ and $z$, such that our function maps any CDM HMF to MDM without the need for a separate $\beta$ value for each ($f, z$) combination.

This is also beneficial from a computational standpoint as the simulations and relevant outputs are not only computationally expensive to run but also require large amounts of memory to store all the raw data (i.\,e.\ snapshots) and post-processed data (i.\,e.\ converted snapshots). If we can ascertain from a limited selection of data a wide range of information about our model then we are able to get results for the HMFs without additional computational resources being needed.

\subsection{Transformation Formula}
 
We model the MDM halo mass function as the CDM halo mass function multiplied by a phenomenological suppression ratio whose characteristic mass scale and shape parameters vary smoothly with the mixed dark matter fraction and redshift. The HMF formula was inspired by work on warm dark matter (WDM) HMF fitting models (e.\,g.\ \citealt{Schneider2012, Lovell2014, Murgia2017}). The WDM models were chosen as a starting point for the fitting model due to the similarities in the phenomenological behaviour of WDM and FDM. WDM also suppresses small-scale structures leading to a \enquote{smoothing} of clumpy structures and producing a similarly diffuse distribution to FDM \citep{Marsh2014}. While the physical origin of the suppression is different in the two models, an analogy can be drawn in the fact that both WDM and FDM suppress structure below a cutoff scale and produce a cutoff in the power spectrum. The parameter \(M_{\rm S}\) is introduced as a phenomenological characteristic mass scale describing the transition in the HMF suppression. Its role is motivated by the existence of characteristic cutoff scales in WDM and FDM cosmologies, although it should not be identified directly with a physical half-mode, Jeans, or de Broglie mass.

We define the suppression ratio as the ratio between the MDM and CDM halo mass functions,
\begin{equation}
    R(M,f,z) \equiv
    \frac{
        \left.\mathrm{d}n/\mathrm{d}\log_{10}M\right|_{\rm MDM}
    }{
        \left.\mathrm{d}n/\mathrm{d}\log_{10}M\right|_{\rm CDM}
    }.
    \label{eqn:ratio definition}
\end{equation}
The MDM halo mass function can therefore be predicted from a CDM halo mass function according to
\begin{equation}
    \left.\frac{\mathrm{d}n}{\mathrm{d}\log_{10}M}\right|_{\rm MDM}
    =
    R(M,f,z)
    \left.\frac{\mathrm{d}n}{\mathrm{d}\log_{10}M}\right|_{\rm CDM}.
\end{equation}
We model this suppression ratio as
\begin{equation}
    R(M,f,z) =
    \left[
        1 +
        \left(
            \frac{M_{\rm S}}{M}
        \right)^{\alpha}
    \right]^{-\beta},
\label{eqn:shortformula} 
\end{equation}
where $M$ is the halo mass in units of $M_\odot/h$, $M_{\mathrm{S}}$ is the suppression mass scale, $\alpha = a_{0} + a_f f + a_z z$, $\beta = b_{0} + b_f f + b_z z$, and $R$ is the suppression ratio. $\alpha$ controls the transition sharpness and $\beta$ controls the suppression strength.

The suppression mass scale is parameterised as
\begin{equation}
\log_{10}\mleft(\frac{M_{\mathrm{S}}}{M_\odot/h}\mright)
= m_0 + m_f f + m_z z \equiv m_{\mathrm{S}}.
\end{equation}
This gives a full model of the form:
\begin{equation}
    R(M,f,z) =\left[1 +\left(\frac{10^{m_{\mathrm{S}}}}{M}\right)^{\alpha} \right]^{-\beta}.
    \label{eqn:fullformula}
\end{equation}

Due to this choice of scaling, in the high-mass regime where $(M_{\mathrm{S}}/M)^{\alpha} \ll 1$ then $R\approx1$ so there is no suppression. In the low-mass regime where $(M_{\mathrm{S}}/M)^{\alpha} \gg 1$ then $R\approx(M/M_{\mathrm{S}})^{\alpha\beta}$ and there is strong suppression.

The parameter $M_{\rm S}$ acts as a phenomenological characteristic mass scale controlling the suppression of the halo mass function. Its inclusion is motivated by the existence of characteristic suppression scales in FDM cosmologies, where wave effects inhibit the growth and collapse of sufficiently small-scale density fluctuations. However, $M_{\rm S}$ should not be interpreted directly as a measurement of a physical scale such as the FDM Jeans, de Broglie, or half-mode mass. The physical quantity of interest is instead the resulting suppression ratio $R(M,f,z)$ over the range in which the model is calibrated.

Best-fitting parameters were obtained by minimising the non-linear least-squares objective with a Levenberg–Marquardt algorithm, starting from an initial parameter vector. 

The fitted parameter vector is
\begin{equation}
    \boldsymbol{\theta}
    =
    (m_0,m_f,m_z,a_0,a_f,a_z,b_0,b_f,b_z).
\end{equation}
The initial parameter vector used was
\begin{equation}
    \boldsymbol{\theta}_0 =
    \left(
        \mathrm{10.0, 0.0, 0.0, 1.0, 0.0, 0.0, 1.0, 0.0, 0.0}
    \right).
\end{equation}
Since the Levenberg-Marquardt algorithm is a local optimiser, we tested the sensitivity of the fit to its initial conditions using a multi-start procedure. We performed 50 tests from different initial parameter configurations including the initial parameter vector with the remaining starting points drawn from broad ranges in each parameter.

In this case the function is the suppression model $R$ and the data to fit are the MDM HMFs. We minimised the $\chi^2$ statistic:
\begin{equation}
    \chi^{2} = \sum_{i}
    \frac{
    [n_{\mathrm{MDM}}(M_{i}) - n_{\mathrm{model}}(M_{i})]^2
    }{\sigma_{i}^2},
    \label{eqn:chisq}
\end{equation}
where $n_{\mathrm{MDM}}(M_i)$ is the measured halo mass function in mass bin $i$, $n_{\mathrm{model}}(M_i)$ is the corresponding model prediction, and $\sigma_i$ is the uncertainty in the measured halo mass function. Assuming Poisson uncertainties in the number of haloes,
$N_i$, in each mass bin, we take
\begin{equation}
    \sigma_i =
    \frac{\sqrt{N_i}}
    {V\,\Delta\log_{10}M_i},
\end{equation}
where $V$ is the simulation volume and $\Delta\log_{10}M_i$ is the logarithmic width of mass bin $i$. This weighting ensures that bins with low halo counts, particularly at the high-mass end, are appropriately downweighted in the fit. The fit corresponds to a Gaussian approximation to the Poisson likelihood. This approximation is expected to be adequate in the well-populated mass bins used for the fit, especially since bins with fewer than five haloes are excluded. A full Poisson likelihood, such as that used in number-count analyses of cosmic voids (e.\,g.\ \citet{London2026}), may provide a more appropriate treatment in poorly populated bins and could improve the modelling of the low-mass end, but we leave this extension for future work.

The model was created from a selection of data at redshifts between $1 \leq z \leq 4$. Ten snapshots were used at redshifts $z = \numlist{4; 3.5; 3; 2.7; 2.2; 2; 1.74; 1.36; 1.15; 1}$ for CDM and MDM where $f = \numlist{0.01; 0.1; 0.2; 0.3}$. We used snapshots from a selection of redshifts within the range that had been rigorously tested in \citet{Dome2025}, while leaving some unsampled redshifts to allow for testing the performance of the fit (see \cref{sec:validatingmodel}). This model is calibrated only for $m = \SI{3.16e-25}{\eV}$, as our simulations only have one axion mass available.

We choose to fit over the range of masses for which the HMF resembles a power law. This reflects the region where the effects of the FDM suppression are clearest. We avoid fitting the low-mass end, and also avoid any noise from the turnover of the HMF curve which is an artefact caused by the grid resolution rather than a physical result (see \cref{fig:halofinderthreshold}). This low-mass cutoff is introduced to exclude halo finder artefacts rather than reflecting a physical suppression scale. For $f > 0$, there is a distinct, physical suppression of the HMF at low masses. Since the turnover may in principle depend on redshift and FDM fraction, we examined its behaviour for a subset of the simulated parameter space. While some variation in the turnover mass is expected, no strong systematic trend was observed in the cases tested. We therefore adopt a conservative fixed lower mass cut of $M = \SI{2e10}{\per\hHubble\Msun}$, chosen to lie above the turnover for the $\rho/\bar{\rho}=60$ case used throughout the analysis. This turnover arises from the finite spatial resolution of the density grid, while its precise location depends on the adopted overdensity threshold. The chosen mass cut therefore excludes the resolution-affected regime while maintaining a consistent fitting procedure across all redshifts and cosmologies.

We also removed any high-mass bins which had fewer than five haloes. This was to avoid any errors on the high-mass end of the fit due to the impact of the shot noise. This is especially important as the grid-based halo finder often identifies very high-mass haloes as one single object instead of multiple high-mass objects and this choice is sensitive to the exact overdensity criterion. This high-mass cut was calculated for each individual HMF in the sample. Thus each has a slightly different fitting region. We use the same logarithmically spaced mass-bin edges for all HMFs, allowing direct comparison between different FDM fractions and redshifts. The binning is defined over the full mass range before applying the low- and high-mass cuts described above. Bins outside the reliable mass range are excluded from the fit, but the remaining data are not rebinned. This ensures that a given mass bin represents the same mass interval in every simulation and avoids introducing differences in the fitted HMFs purely through changes in binning.

Since the absolute HMF depends on the overdensity criterion used by the grid-based halo finder, we test how this systematic propagates into the suppression ratio. We repeated halo finding using overdensity criteria $\rho/\bar{\rho} = 30$ and $120$ and compared it with the fiducial $\rho/\bar{\rho} = 60$ results. 

We define the change relative to the fiducial threshold as
\begin{equation}
    \Delta_{\rm halo}
    =
    \log_{10}
    \left(
    \frac{R_{\Delta}}{R_{60}}
    \right),
\end{equation}
where $R_{\Delta}$ is the MDM-to-CDM HMF suppression ratio measured using overdensity threshold $\Delta=\rho/\bar{\rho}$. We quantify the size of the systematic using the median absolute offset, $\mathrm{median}(|\Delta_{\rm halo}|)$, over the common mass bins.

At $z=1$, the median absolute offsets are $0.061\,\mathrm{dex}$ and $0.019\,\mathrm{dex}$ respectively. At $z=4$, the corresponding offsets are $0.195\,\mathrm{dex}$ and $0.064\,\mathrm{dex}$. Thus, we found that much of the halo definition dependence cancels when forming the MDM-CDM ratio. However, a residual systematic remains, particularly for lower overdensity criterion at high redshift. This represents a deliberately conservative test of the halo definition as a substantially lower threshold allows more diffuse density structures to be linked into individual haloes. In addition, $z=4$ lies at the upper edge of the redshift range calibrated in this work, where the halo population is less well sampled and the measured HMF is correspondingly more sensitive to halo-finder choices. The larger offset in this case therefore provides a useful estimate of the systematic uncertainty in an extreme combination of halo-finder criterion and redshift, rather than being representative of the typical behaviour across the calibrated range.

\subsection{Success of model fit}

\begin{figure*}
    \centering
	\includegraphics[width=\textwidth]{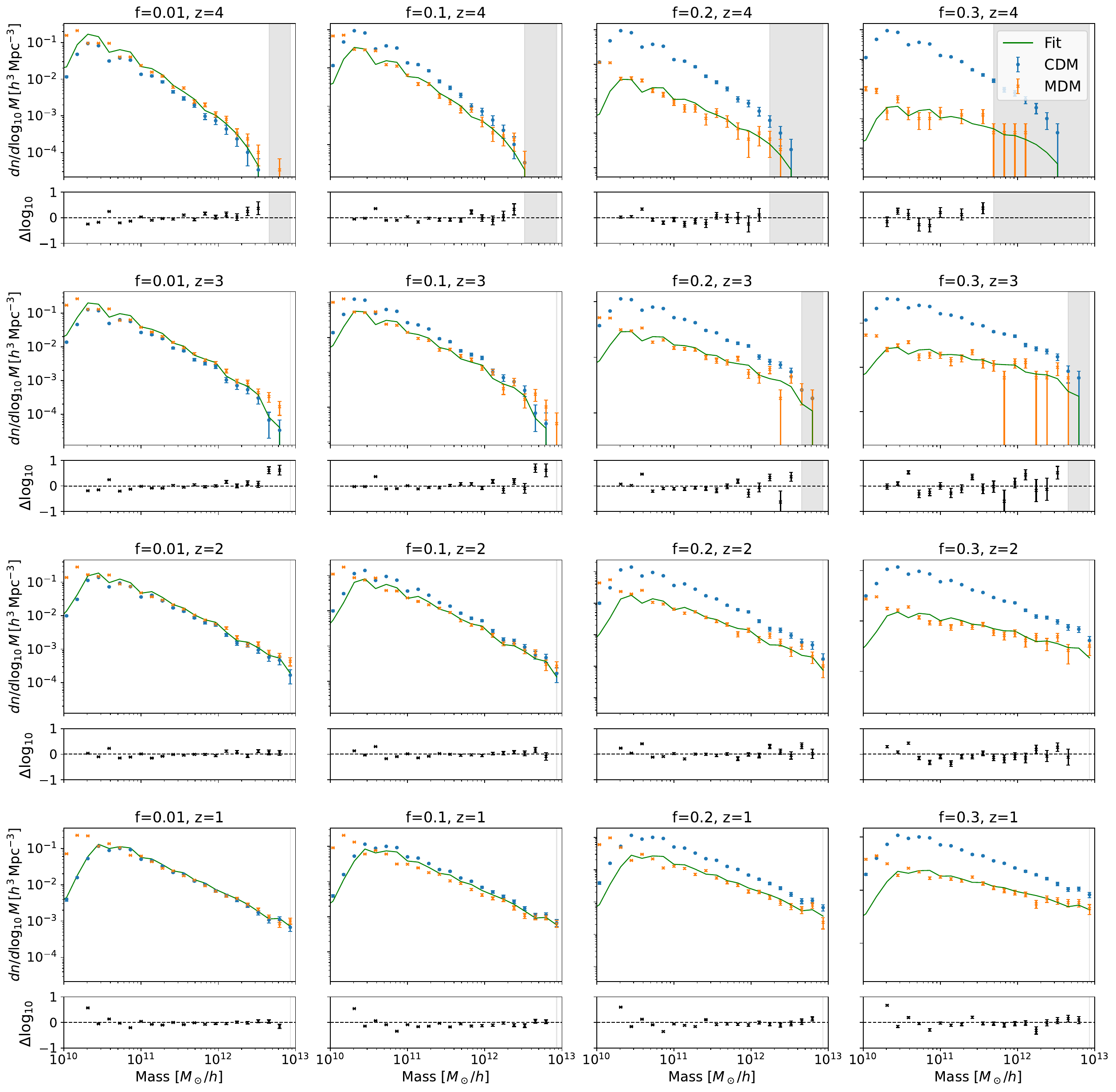}
    \caption{Examples show the performance of the fit for four different redshift values for the four different FDM fractions, as labelled on each panel. CDM points are shown in blue dots on each plot for reference, with MDM points shown as orange crosses. The grey shaded region represents $M_{200\mathrm{c}}$ halo mass ranges excluded from the fits. The lower mass bins on the left-hand side are excluded due to the turnover in the HMF as this is an artefact of the halo finder rather than a physical effect. At the high-mass end there is a threshold cut applied and bins containing fewer than five haloes are excluded, as these numbers are too low to effectively use for the HMF matching. The green line shows the predicted MDM fit from the transformation of the CDM model. Error bars show HMF uncertainties propagated from Poisson halo-count uncertainties.}
    \label{fig:HMFfit}
\end{figure*}

The results of this fitting procedure can be seen in \cref{fig:HMFfit}. Our predicted MDM model (shown in green) traces the simulated MDM points (shown in orange) well. The best fitting parameters and uncertainties are shown in \cref{tab:fitvalues}. 

Across the fitted mass range, the model typically agrees with the simulated HMF to within $\leq 0.1\,\mathrm{dex}$ ($\simeq \SI{25}{\percent}$) for low FDM fractions and $\leq 0.2$\,dex ($\simeq \SI{60}{\percent}$) more generally, as shown by the residuals in the lower panels. Larger deviations occur at the high-mass end and for higher FDM fractions, where Poisson noise dominates.

At lower FDM fractions the CDM and MDM results become increasingly similar, making the differences more difficult to distinguish. However, at higher redshifts and higher FDM fractions the difference between the models and the predictions of the fit are clearer.

\begin{table}
\centering
\caption{Best fitting parameters and uncertainties for the model described in \cref{eqn:fullformula}. The full model was fitted using all fractions $f = \numlist{0.01; 0.1; 0.2; 0.3}$ whereas the limited model uses only $f=\numlist{0.1;0.2;0.3}$. Parameter uncertainties were estimated from the covariance matrix returned by the fitting procedure.}
\label{tab:fitvalues}

\begin{tabular}{
c
S[table-format=+2.2(3), separate-uncertainty=true]
S[table-format=+2.2(3), separate-uncertainty=true]
}
\toprule
{Parameter} & {Full Model} & {Limited Model} \\
\midrule
$m_0$ & 15.18 +- 0.37 & 14.82 +- 0.48\\
$m_f$ & -0.76 +- 1.16 & 0.91 +- 1.72\\
$m_z$ & -0.23 +- 0.09 & -0.29 +- 0.10\\
$a_0$ & -0.58 +- 0.94 & 0.06 +- 0.59\\
$a_f$ & 8.63 +- 0.14 & 4.53 +- 0.45\\
$a_z$ & 1.30 +- 0.21 & 1.04 +- 0.10\\
$b_0$ & -0.02 +- 0.31 & -0.03 +- 0.30\\
$b_f$ & 0.38 +- 0.63 & 0.51 +- 0.51\\
$b_z$ & 0.00 +- 0.06 & 0.00 +- 0.17\\
\bottomrule
\end{tabular}
\end{table}

The predicted MDM model reproduces the simulated HMFs well over the well-populated mass range, particularly for low FDM fractions. The largest deviations occur at the high-mass end and for higher FDM fractions, especially at higher redshift. These bins contain relatively few haloes and therefore have large Poisson uncertainties. As a result, they carry less statistical weight in the fit than the well-populated intermediate-mass bins, even when the fractional difference between the model and the simulation appears large. The agreement improves towards lower redshift as more haloes form and the HMF becomes better sampled. Additionally, we do not explicitly model cosmic variance, but it is possible that the finite simulation volume may introduce additional uncertainty, particularly at the high-mass end where halo counts are low. Consequently, some of the apparent bin-to-bin fluctuations in the measured HMFs should not be over-interpreted, and the fit should be regarded as a phenomenological description of the broad suppression
trend rather than a precise model of every local feature in the binned HMF.

To assess the stability of the fitted parameters, we repeated the optimisation from multiple initial parameter combinations. We find that substantially different combinations of the individual coefficients can produce effectively indistinguishable predictions over the calibrated mass range. This confirms that the fitting function is strongly parameter-degenerate.

This behaviour arises because the characteristic mass scale lies above the masses directly sampled by the simulations, such that the fitting function is predominantly evaluated in its asymptotic regime, 
\begin{equation}
    R \simeq
    \left(\frac{M}{M_{\rm S}}\right)^{\alpha\beta}.
\end{equation}
Consequently, combinations of $M_{\rm S}$, $\alpha$, and $\beta$ are much more strongly constrained than the individual parameters themselves. We therefore do not interpret the fitted coefficients as independently measured physical quantities. The relevant result is the predicted MDM-to-CDM HMF ratio within the calibrated range of mass, FDM fraction, and redshift.

The fit could potentially be improved by a model with a curved suppression or a scale-dependent turnover, but as we will discuss later experimenting with these parameters did not have a notable impact on the fit.

One particular difficulty we faced was in tuning the fits for small FDM fractions. As the effect of the FDM is not very significant in small amounts, the $f_{\mathrm{FDM}} = 0.01$ and $f_{\mathrm{FDM}} = 0.1$ results are very similar, and both are similar to the CDM-only case. A larger range
of FDM fractions could be used to further fine-tune the model. However, given the constraints the model provides an acceptable fit.

Some of the specific slope characteristics of the MDM model predicted by the simulation are also captured by the fitting formula in \cref{eqn:fullformula}, e.\,g.\ the change in slope between the CDM and MDM at $f_{\mathrm{FDM}} = 0.2$. The model therefore captures changes in HMF slope rather than only a mass-independent shift. This can be seen in the residuals on the plots in \cref{fig:HMFfit}.

 \begin{figure*}
    \centering
	\includegraphics[width=\textwidth]{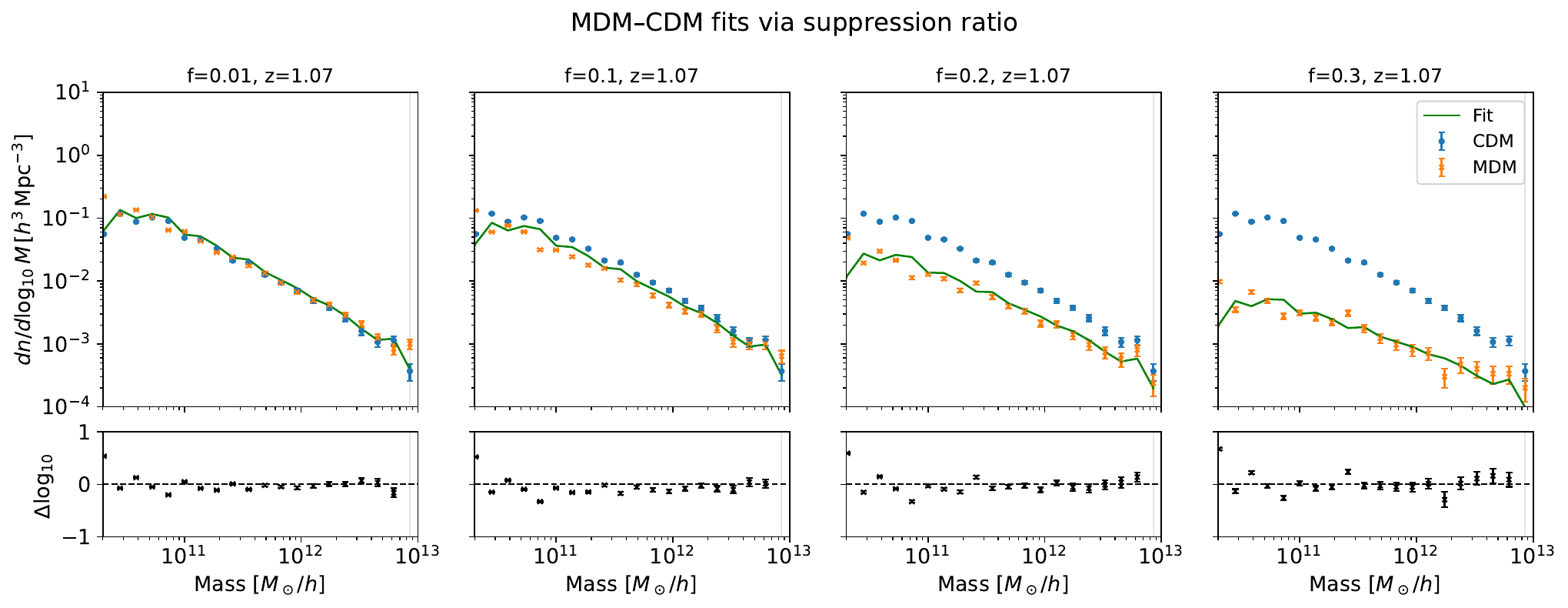}
    \caption{Model predictions for data not used in the initial fit. The CDM HMF at $z = 1.07$ is shown as blue circles. The actual MDM results are shown as orange crosses and the green fit line shows the predicted HMF shape. The bottom panels show the residuals. All points have Poisson errors. The model fit can be seen to predict the MDM HMF well at small fractions but diverges more at higher fractions. This is similar to the trend seen in the data used to produce the fit.}
    \label{fig:HMFprediction}
\end{figure*}

One limitation of the fitting method is that it is calibrated only above the adopted lower mass cut and therefore does not model the behaviour of the HMF in the low-mass regime. This choice is intentional, as the low-mass end of the measured HMF is affected by the resolution turnover of the grid-based halo finder and does not provide a reliable constraint on the fit. Consequently, the fitting function captures the physical suppression of halo abundances above the resolution limit but does not attempt to reproduce either the numerical turnover present in the measured HMFs or any physical behaviour below the calibrated mass range. Thus this fitting model should only be trusted for masses above the lower mass threshold for the fit, and not used to extrapolate to lower masses.

We also experimented with whether adding a high-mass floor to account for the non-linearity of the MDM curves at higher fractions improved the model fit. We found that the curve fitting always fits the floor model with a factor very close to zero meaning the contribution of the floor was always essentially erased. This could be because the non-linearity is a feature of higher FDM fraction models and therefore the floor needs to be fitted to take into consideration both FDM fraction and mass. However, even when accounting for this we could not find a model that provided a significantly better fit. Also due to the low numbers of haloes present in the higher mass bins it is difficult to tell how much of a floor fitting model would be impacted by the highly variable slope and how much of this variation is dominated by errors. Thus we removed the floor from the model and allowed the fit to be done as if the HMF slope was perfectly linear.

\subsection{Comparison with existing halo mass function predictions}
\label{sec:hmf_comparison}
To place the empirical mapping introduced in this work in the context of existing predictions, we compare the measured MDM-to-CDM HMF ratios with the prescription of \citet{Dome2025}, and with classical Press-Schechter (PS; \citealt{PressSchechter1974}) and Sheth-Tormen (ST; \citealt{ShethTormen1999}) calculations.
    
For the PS and ST calculations, we use matched MDM and $\Lambda$CDM linear matter power spectra generated with \texttt{AxionCAMB} \citep{Hlozek2015}. For each FDM fraction, the MDM and $\Lambda$CDM spectra are generated using the same cosmological parameters and numerical setup. The linear variance is calculated using a spherical top-hat filter, 
\begin{equation}
        \sigma^2(M,z)
        =
        \frac{1}{2\pi^2}
        \int
        P_{\rm L}(k,z)
        W^2(kR)
        k^2\,{\rm d}k,
\end{equation}
where
\begin{equation}
        R =
        \left(
            \frac{3M}{4\pi\bar{\rho}_{\rm m}}
        \right)^{1/3},
        \qquad
        W(x)=3\frac{\sin x-x\cos x}{x^3},
\end{equation}
and the peak height is $\nu=\delta_{\rm c}/\sigma$, with
    $\delta_{\rm c}=1.686$.
    
For PS, the multiplicity function is
\begin{equation}
        f_{\rm PS}(\nu)
        =
        \sqrt{\frac{2}{\pi}}\,
        \nu
        \exp\left(-\frac{\nu^2}{2}\right),
\end{equation}
while for ST we use
\begin{equation}
        f_{\rm ST}(\nu)
        =
        A\sqrt{\frac{2q}{\pi}}\,
        \nu
        \left[1+(q\nu^2)^{-p}\right]
        \exp\left(-\frac{q\nu^2}{2}\right),
\end{equation}
with $A=0.3222$, $p=0.3$, and $q=0.707$, following \citet{Dome2025}. The corresponding HMF is
\begin{equation}
        \frac{{\rm d}n}{{\rm d}\log_{10} M}
    =
    \frac{\ln 10}{2}
    \frac{\bar{\rho}_{\rm m}}{M}
    f(\nu)
    \left|
        \frac{{\rm d}\ln\sigma^2}
             {{\rm d}\ln M}
    \right|.
\end{equation}
    
We compare the models using the same MDM-to-CDM HMF ratio calibrated in this work (Eq.~\ref{eqn:ratio definition}), rather than the absolute HMF. For PS and ST, the numerator and denominator are evaluated independently using the matched MDM and $\Lambda$CDM linear power spectra. This also reduces sensitivity to differences in HMF normalisation and halo definition between the different approaches. The PS/ST calculation provides a physical connection between the scale-dependent suppression of the linear power spectrum and the resulting change in halo abundance. To quantify the agreement with the  simulations, we define the logarithmic residual
\begin{equation}
        \Delta
        =
        \log_{10}R_{\rm model}
        -
        \log_{10}R_{\rm sim},
\end{equation}
and calculate the root-mean-square (RMS) error and median absolute residual over the resolved mass bins used in the calibration.
    
\begin{figure*}
        \centering
        \includegraphics[width=\textwidth]
            {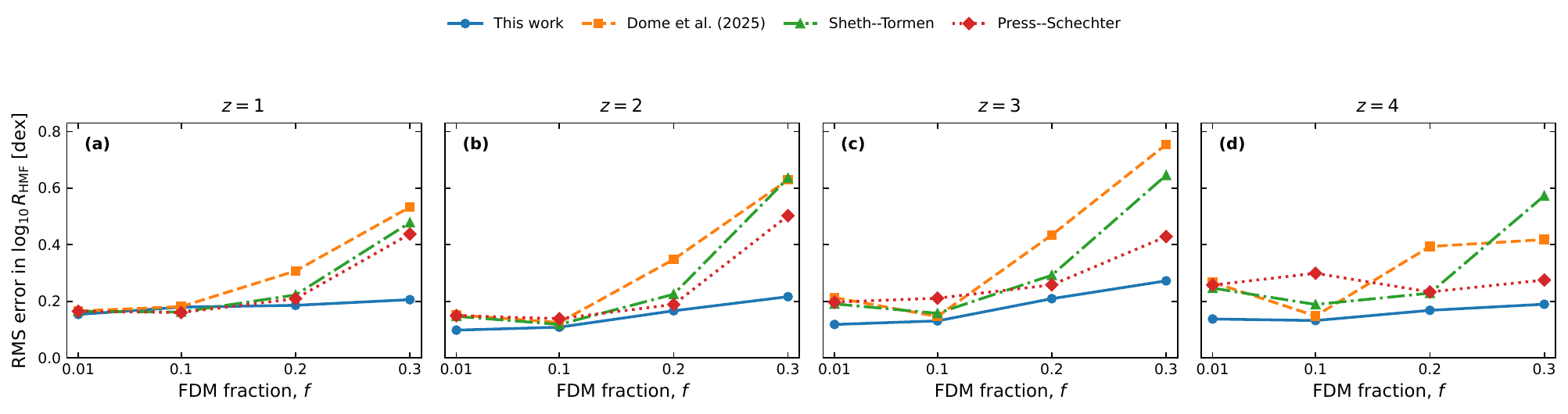}
        \caption{
        Root-mean-square (RMS) error between the predicted and measured MDM-to-CDM HMF ratios as a function of FDM fraction at $z=1$, $2$, $3$, and $4$. Results are shown for the empirical mapping introduced in this work, the \citet{Dome2025} prescription, and PS and ST predictions calculated using the corresponding linear \texttt{AxionCAMB} matter power spectra. The relative performance of the prescriptions varies with both FDM fraction and redshift, with the largest differences generally occurring towards higher FDM fractions. Fewer resolved mass bins are available at the largest fractions and redshifts, and these points should therefore be interpreted with greater caution.
        }
        \label{fig:hmf_model_rmse}
\end{figure*}
    
Figure~\ref{fig:hmf_model_rmse} shows the RMS error as a function of FDM fraction at each redshift. The empirical mapping generally gives the smallest residuals, with the improvement over the other prescriptions becoming more pronounced towards larger FDM fractions. The dependence on redshift is less systematic, although the relative performance of the models changes across the redshift range.
    
\begin{table}
        \centering
        \caption{
        Comparison between the measured MDM-to-CDM HMF ratios and the     different HMF prescriptions over the complete calibration dataset. Residuals are defined as
        $\Delta=\log_{10}R_{\rm model}-\log_{10}R_{\rm sim}$.
        }
        \label{tab:hmf_model_comparison}
        \begin{tabular}{lcc}
            \hline
            Model
            & RMS error [dex]
            & Median $|\Delta|$ [dex] \\
            \hline
            This work          & 0.169 & 0.092 \\
            Press--Schechter   & 0.272 & 0.140 \\
            Sheth--Tormen      & 0.313 & 0.146 \\
            \citet{Dome2025}   & 0.356 & 0.188 \\
            \hline
        \end{tabular}
\end{table}
    
Across the complete calibration dataset, our empirical mapping gives an RMS error of $0.169$ dex, compared with $0.272$ dex for PS, $0.313$ dex for ST, and $0.356$ dex for the \citet{Dome2025} prescription. The corresponding median absolute residual is also smallest for the empirical mapping (Table~\ref{tab:hmf_model_comparison}). This demonstrates that the additional dependence on FDM fraction and redshift in our calibration provides a closer description of the HMF ratios measured in these simulations, particularly towards larger FDM fractions.
    
We note that \texttt{AxionHMcode} is primarily a halo-model prediction for the nonlinear matter power spectrum rather than an independent empirical fitting function for the total MDM HMF, and we therefore do not present a separate ``AxionHMcode HMF'' comparison.
    
The smaller residuals of the empirical mapping should not be interpreted as demonstrating universal superiority over existing HMF prescriptions. The model was calibrated directly to the HMF measurements considered here, whereas the PS, ST, and \citet{Dome2025} prescriptions were not recalibrated to these halo catalogues. The comparison therefore demonstrates improved descriptive accuracy within the calibrated simulation set. Predictive robustness is assessed separately using the validation tests described in Section~\ref{sec:validatingmodel}.

\subsection{Validating the model fit}
\label{sec:validatingmodel}

\Cref{fig:HMFprediction} shows the model fit used on snapshot data not included in the initial fit. This uses the CDM HMF from $z = 1.07$ as input. The model reproduces the overall suppression of the MDM HMF across the tested FDM fractions, with residuals comparable to those obtained for the snapshots used in the calibration. The agreement is particularly good for the lower FDM fractions, while the largest residuals occur for the highest fractions and in mass bins where the simulated HMF is noisier. We do not expect the model to reproduce every fluctuation in the simulated HMF, since these features are sensitive to low halo counts and halo finder specifics. Instead, the aim is to capture the smooth, systematic dependence of the HMF on $f$ and $z$, which the validation test shows is achieved to good accuracy.

\begin{figure*}
    \centering
	\includegraphics[width=\textwidth]{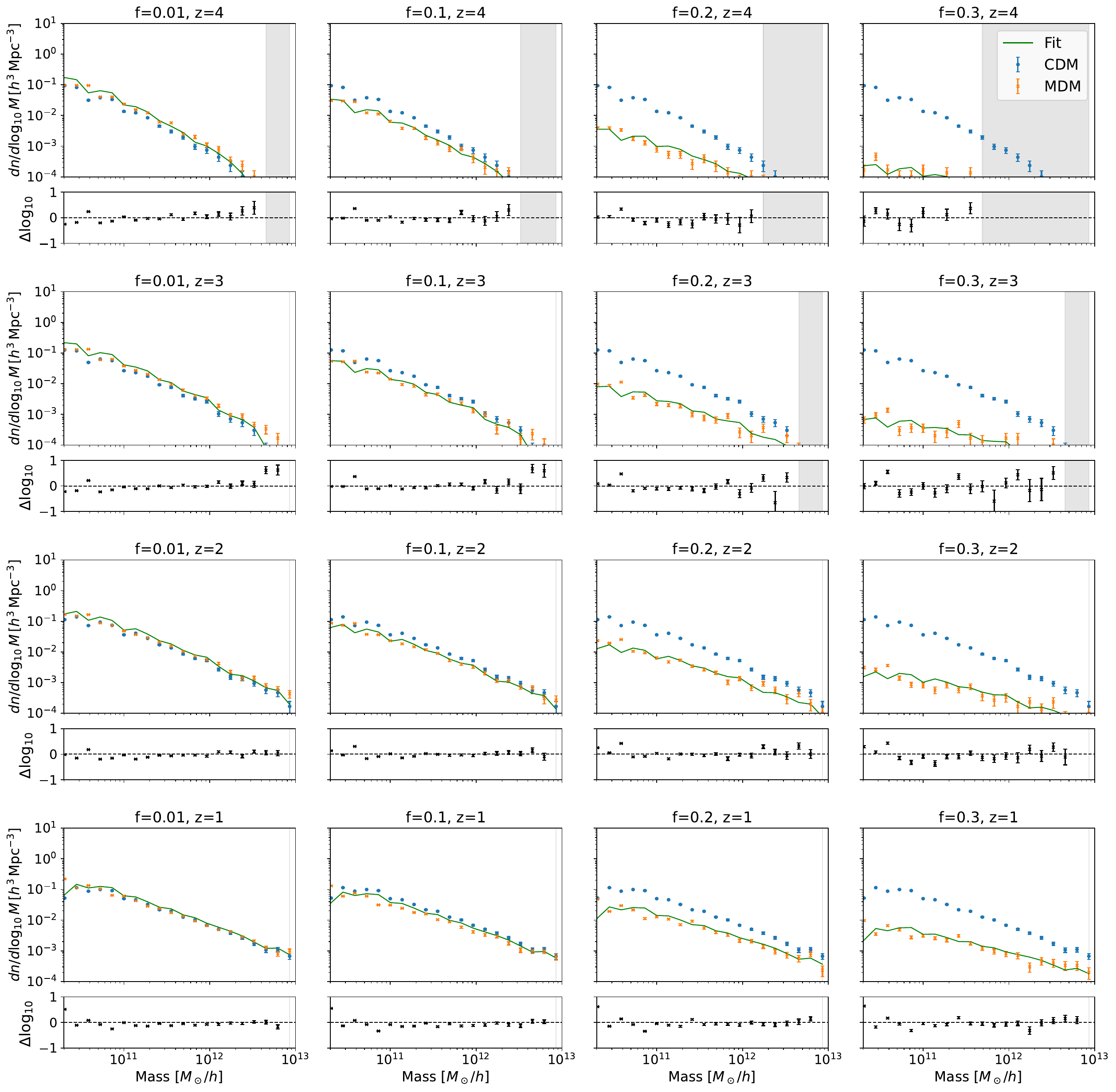}
    \caption{Examples show the performance of the fit for four different redshift values for four different FDM fractions using a model fit from only $f = \numlist{0.1; 0.2; 0.3}$. CDM-only cosmology points are shown in blue dots on each plot for reference with MDM points shown as orange crosses. The grey shaded region represents halo mass ranges excluded from the fits. The lower mass bins on the left hand side are excluded due to the turnover in the HMF as this is an artefact of the halo finder rather than a physical effect. At the high-mass end there is a threshold cut applied that any bin with fewer than five haloes in is excluded as these numbers are too low to effectively use for the HMF matching. The green line shows the predicted MDM fit from the transformation of the CDM model. Error bars show HMF uncertainties propagated from Poisson halo-count uncertainties.}
    \label{fig:HMFfit_limit}
\end{figure*}

\begin{figure*}
    \centering
	\includegraphics[width=\textwidth]{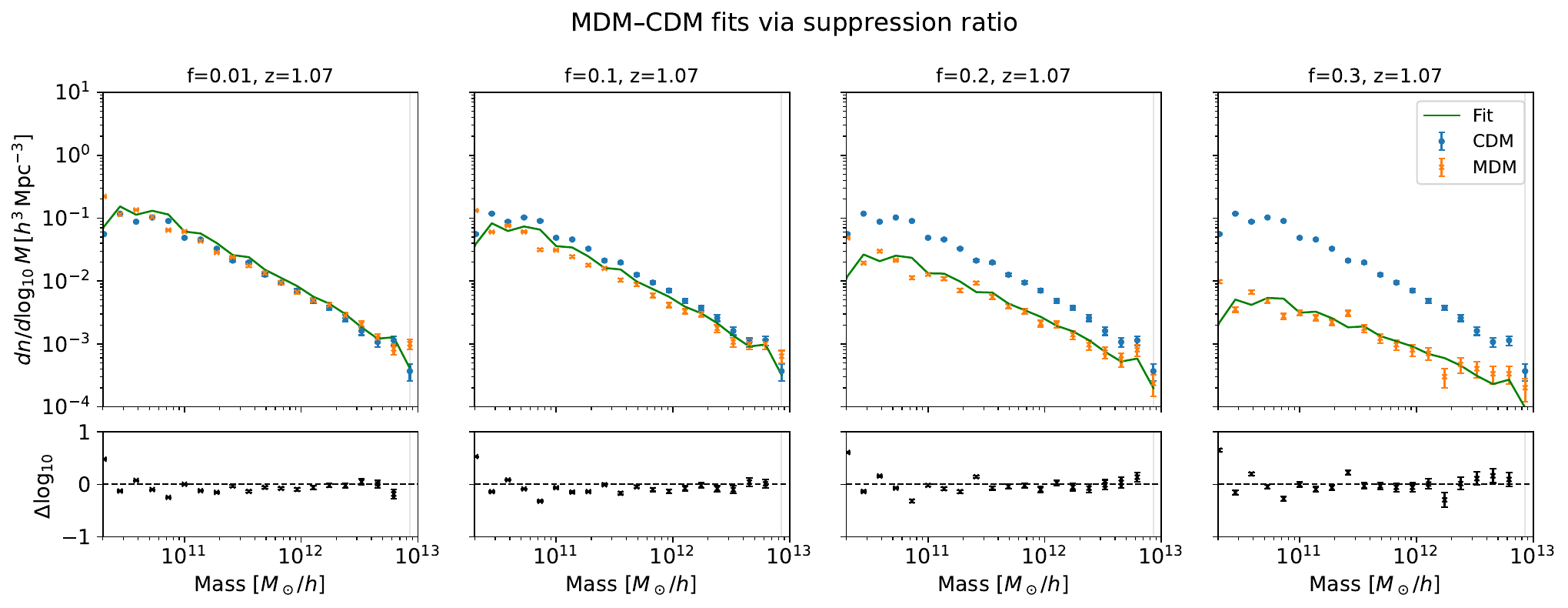}
    \caption{Model predictions for data not used in the initial fit for the limited model using only $f = \numlist{0.1; 0.2; 0.3}$. The CDM HMF from $z = 1.07$ is shown as blue circles. The actual MDM results are shown as orange crosses and the green fit line shows the predicted HMF shape. The bottom panels show the residuals. All points have Poisson errors. The limited model fit can be seen to be very similar to that of the full model, showing the low fractions are not skewing the model fit.}
    \label{fig:HMFprediction_limit}
\end{figure*}

To test whether the fit could be improved, we investigate if the model should be fitted where the impact of the FDM is more obvious i.\,e.\ where the difference in the HMF is largest. For $f = \numrange{0.01}{0.1}$, the HMF is very similar to that of pure CDM. To investigate whether our model fit was being disrupted by this similarity we also fitted a second \enquote{limited model} which does not use the $f=0.01$ data in the fit. The data used for the fitting is shown in \cref{fig:HMFfit_limit} and the results for the parameters determined from this limited model are also listed in \cref{tab:fitvalues}. \Cref{fig:HMFfit_limit} shows that removing the $f=0.01$ data does not significantly improve the high fraction fits, but does cause the low fraction fit to be over-predicted. This shows that the low fraction data needs to be included in the model to correctly predict across the whole fraction range, and the high fraction fits are not improved by just fitting the model where the deviations between the CDM and MDM are more obvious. The prediction from the limited model is shown in \cref{fig:HMFprediction_limit}. The limited model performs worse in making a prediction for the MDM HMFs, showing that using the full model which included all the MDM fractions is better for making predictions.

We also considered whether the model should be fitted in two regimes i.\,e.\ one fit for the low fractions ($0 \leq f < 0.1$) and another fit for the higher fractions ($0.1 \leq f \leq 0.3$). Our limited model could be used as the higher fraction fit and the low fraction fit could be given to be equivalent to the CDM model. However, given our limited selection of data we have no test cases other than our $f = 0.01$ model to fit this lower-end model on and so cannot make any predictions on this.

Overall this leads us to use our full model fit as the best available option for modelling MDM across a range of redshifts and fractions.

\subsection{Future Improvements}
A fitting method such as this would benefit from additional data points. In particular, simulations at more closely spaced FDM fractions, for example between $f=0.1$ and $f=0.2$, would allow a more detailed investigation of the range of FDM fractions over which the HMF begins to depart significantly from the CDM prediction. Extending the simulations to substantially larger fractions, $f>0.3$, is likely to be less beneficial, as these models are increasingly strongly constrained by current observations and would require more computationally expensive simulations for comparatively limited scientific gain.

Additional redshift sampling would also improve the description of how the HMF evolves with time. At lower redshifts, the existing simulations provide relatively dense sampling, whereas at higher redshifts the model is limited by the number of available simulation snapshots. Additional outputs or simulations targeting this regime would therefore allow the redshift dependence of the fit to be constrained more accurately.

The FDM particle mass could also be included as an additional parameter in the fit. The ultralight axion regime encompasses a broad range of particle masses, and extending the simulation suite to different values of $m_{\mathrm{FDM}}$ would allow the dependence of the HMF suppression on both FDM fraction and particle mass to be investigated.

With a larger suite of simulations, the dependence of the HMF suppression on the model parameters could instead be represented using a statistical emulator. For example, a Gaussian-process or other surrogate model could be trained on the measured suppression ratio, $R(M,f,z)$, to interpolate between the simulated values of FDM fraction and redshift, providing predictions at parameter combinations that have not been simulated directly. Extending the simulation suite to different FDM particle masses would allow $m_{\mathrm{FDM}}$ to be included as an additional emulator parameter.

\section{Discussion}
\label{sec:Discussion}

\subsection{Physical interpretation of MDM structure formation suppression}
Our results demonstrate that the primary effect of introducing an FDM component into a CDM cosmology is a suppression of small-scale structure, while leaving large-scale clustering largely unchanged. This behaviour is consistent with expectations from the linear matter power spectrum, where the finite de Broglie wavelength of ultralight axions introduces a characteristic cutoff below which density fluctuations are suppressed. In the non-linear regime probed by our simulations, this appears as a reduction in the abundance of low-mass haloes and a delay in their formation.

The similarity in large-scale structure across all FDM fractions reflects the fact that FDM traces the dominant CDM component. Consequently, the locations of the most massive haloes remain largely unchanged, even at higher FDM fractions. However, the internal structure and surrounding environments of these haloes differ, with higher fractions producing smoother density fields and less pronounced substructure. This is clearly visible in the large-scale density maps (\cref{fig:cosmicwebs}), where increasing FDM fraction leads to more diffuse filaments and larger void regions.

The redshift evolution of the HMF in MDM models broadly follows the hierarchical growth expected in CDM cosmologies, with haloes accreting mass over time and shifting towards higher mass bins. However, the presence of FDM delays the formation of low-mass progenitors, leading to a reduced population of haloes at early times. This effect becomes more pronounced at higher redshifts and higher FDM fractions, where the suppression of small-scale fluctuations is strongest.

We also observe a change in the slope of the HMF at higher masses for larger FDM fractions. While part of this effect may be physical, reflecting differences in merger histories or accretion rates, it is likely also influenced by the halo-finding methodology. The grid-based overdensity criterion may preferentially merge nearby structures into single high-mass haloes. This can particularly affect cosmologies with higher FDM fractions as the haloes are more diffuse and thus have less clearly defined \enquote{edges}, so it is easier for multiple haloes to be grouped into a single object rather than located separately. This represents a phenomenological departure from the CDM cases, which can be exacerbated by the grid-based halo finder. This shows the importance of carefully interpreting high-mass behaviour in MDM simulations.

\subsection{Observational Implications}
The suppression of small-scale structure in MDM models is qualitatively similar to that seen in WDM scenarios and in CDM models with strong baryonic feedback. This raises the possibility of degeneracies between these physical effects when interpreting observational data. While WDM and FDM models directly suppress the formation of low-mass dark matter haloes, baryonic processes act to suppress star formation within such haloes, thereby reducing the number of observable, luminous systems. Additionally, both FDM and baryonic feedback can lead to smoother central density profiles, making it challenging to differentiate their contributions using current observations.

Future surveys with improved sensitivity to high-redshift galaxies and small-scale structure may help break these degeneracies. Measurements of the abundance of faint galaxies, the properties of dwarf galaxies, and the detailed structure of the cosmic web could provide constraints on the allowed FDM fraction. 

Strong gravitational lensing provides a promising observational probe of low-mass dark matter structure, with current and future analyses sensitive to haloes with masses of approximately $10^{7}$--$10^{9}\,\mathrm{M_\odot}$. This mass range lies below the range over which the HMF transformation presented in this work is calibrated, which for $m=10^{-24.5}\,\mathrm{eV}$ is approximately
$10^{11}$--$10^{13}\,h^{-1}\mathrm{M_\odot}$. Our fitting function
therefore cannot be applied directly to strong-lensing substructure
constraints without extrapolation. Nevertheless, the suppression of
low-mass halo formation produced by an FDM component is expected to
extend towards the lower halo masses probed by strong lensing.
Simulations with a larger dynamic range would be required to connect
our measured HMF suppression directly to these observations. Such an
extension would allow strong-lensing measurements of low-mass halo
abundances to provide an observational test of the MDM models explored here. However, given the relatively subtle effects for $f \lesssim 0.1$, distinguishing MDM from CDM may remain observationally challenging.

\subsection{Future Work}
Our analysis is subject to several limitations. The simulations are restricted to a single axion mass and a limited range of FDM fractions ($f \leq 0.3$). Future work could address these limitations by exploring a wider range of axion masses and improving the simulation resolution. Extending the HMF transformation model to include additional parameters, such as axion mass or environmental dependence, would also be valuable. Finally, incorporating baryonic physics would allow for a more direct comparison with observations and help assess degeneracies between different physical processes.

Overall, our results demonstrate that MDM models provide a flexible and viable extension to the standard CDM paradigm, with distinct but subtle signatures in the nonlinear structure of the Universe.

\section{Conclusions}
\label{sec:Conclusions} 
We have investigated halo mass functions (HMFs) in mixed dark matter (MDM) cosmologies. To achieve this, we developed a grid-based halo-finding pipeline within the \texttt{AxiREPO} framework, designed to identify haloes in simulations containing both cold dark matter (CDM) and fuzzy dark matter (FDM). By smoothing all components onto a common grid and identifying overdensities, this method enables a unified and physically consistent treatment of both species.

We demonstrate that this grid-based halo finder produces stable and physically meaningful halo catalogues for MDM simulations. Unlike previous approaches that approximate one component as another to enable single-species halo finding, our method treats both components self-consistently, providing a more robust framework for analysing mixed dark matter scenarios.

Our results show that FDM traces the underlying CDM distribution, leading to similar large-scale structure across all models. However, increasing the FDM fraction suppresses small-scale structure due to wave interference effects associated with the large de Broglie wavelength of FDM particles. This suppression reduces the abundance of low-mass haloes and effectively delays structure formation. Correspondingly, we find that the HMF evolves systematically with both FDM fraction and redshift, exhibiting an overall downward shift and changes in slope at higher FDM fractions.

We further derive a transformation formula that accurately maps CDM HMFs to their MDM counterparts across the range of FDM fractions and redshifts explored in this work ($f = \numrange{0.01}{0.3}$ and $z = \numrange{1}{4}$). This model enables efficient prediction of MDM HMFs within this parameter space without requiring a full suite of simulations. In turn, these HMFs can be used to constrain models of small-scale power suppression and identify regions of MDM parameter space that warrant further investigation with dedicated simulations. The model validity is constrained by the reliability of the underlying HMF measurements: at low masses by resolution limits and halo finder artefacts, and at high masses by low-number statistics, finite-volume effects, and potential halo finder systematics. These limitations should therefore be considered when extrapolating beyond the mass range directly constrained by the simulations.

Our results are constrained by the limited FDM fraction range ($f \leq 0.3$) and the resolution of the simulations. We focus on relatively small FDM fractions because larger values generally produce stronger suppression of small-scale structure and are increasingly disfavoured by existing observational constraints on ultralight dark matter. The range considered here therefore targets the currently viable regime in which FDM may remain cosmologically relevant while preserving consistency with observations. Nevertheless, these models provide a useful framework for further exploration of MDM cosmologies.

Future work could extend this analysis to higher FDM fractions, bridging the gap between partial and fully FDM-dominated cosmologies. The fitting model could also be improved through the inclusion of non-linear terms or machine learning-based interpolation to better capture subtle features in the HMF. Additionally, combining the HMF framework with forward models of the matter power spectrum and galaxy formation represents a promising direction.

Overall, mixed dark matter remains a viable extension of the standard CDM paradigm. The methods developed here provide a scalable approach to exploring MDM cosmologies, reducing computational cost while enabling broader and more flexible investigation of dark matter models.

\section*{Acknowledgements}
The authors would like to thank Alex Laguë for useful discussions on this work. We thank the anonymous referee for their helpful comments and suggestions, which improved the clarity of the manuscript. SJ acknowledges a studentship funded in part by Dell Technologies and Durham University. SB is supported by the UKRI Future Leaders Fellowship [grant numbers MR/V023381/1 and UKRI2044]. The authors acknowledge support from the Science and Technology Facilities Council through ST/X001075/1. 
This work used the DiRAC@Durham facility managed by the Institute for Computational Cosmology on behalf of the STFC DiRAC HPC Facility (\texttt{www.dirac.ac.uk}). The equipment was funded by BEIS capital funding via STFC capital grants ST/K00042X/1, ST/P002293/1, ST/R002371/1 and ST/S002502/1, Durham University and STFC operations grant ST/R000832/1. DiRAC is part of the National e-Infrastructure.

\section*{Data Availability}
 
The MDM snapshot data and post-processing scripts are made available upon reasonable request to the lead author. 

A lightweight Python implementation of the calibrated suppression model described by Equations 10 - 14, including the fitted coefficients and example usage, is publicly available at https://github.com/sarahcjohnston/MDM-HMF


\renewcommand{\refname}{REFERENCES}
\bibliographystyle{mnras}
\bibliography{example} 


\appendix
\section{Grid Construction}
The combined CDM and FDM grid is achieved by using the post-processing tools in \texttt{AxiREPO}, in particular the \texttt{OUTPUT\_DENSITY\_GRID} option. The density grid post-processing involves taking the particle distribution from the CDM part of the snapshot and smoothing it onto a grid using a cloud-in-cell (CIC) method. To make the summation of the CDM and FDM components as simple as possible and to avoid introducing additional interpolation for the FDM component, the density grid resolution for the CDM component was chosen to be the same as for the underlying FDM grid, which in this case was of dimension $2048^3$. This resolution is sufficient for our purposes as it allows for low-mass haloes to be detected. Furthermore, as we are focusing on the differences between CDM and MDM at a fixed resolution, as long as the resolution is the same for all our analyses then we can still make direct comparisons. \texttt{AxiREPO} then runs on a simulation snapshot, reads in the particles, applies the mass assignment, and outputs the component as a new \enquote{Density Grid} field in a converted version of the snapshot file.

This procedure produces the original snapshot data with the FDM $\psi$ grid and a new, converted snapshot with the CDM now also as a grid (\cref{fig:density grids}, left and centre). However, to effectively run a halo finder these grids must then be combined. We require the resulting grid to have the form of our normal FDM grid, with real ($\psi_{\mathrm{Re}}$) and imaginary ($\psi_{\mathrm{Im}}$) components, so that the halo finding tools in \texttt{AxiREPO} (described in the following subsection) still work. To do this we run the two snapshots through a processing script which adds the two components together according to the formulae:
\begin{align}
    \psi_{\mathrm{Re}_{\mathrm{Out}}} &= \sqrt{\rho_{\mathrm{CDM}} + \mleft(\psi_{\mathrm{Re}_{\mathrm{FDM}}}^2 + \psi_{\mathrm{Im}_{\mathrm{FDM}}}^2\mright)} ,
    \intertext{and}
    \psi_{\mathrm{Im}_{\mathrm{Out}}} &= 0 .
\end{align}
This results in one final snapshot where the absolute value squared of the \enquote{artificially constructed} FDM grid now represents the total density of all types of dark matter (\cref{fig:density grids}, right).\footnote{%
    It should be noted that the FDM grid constructed in this way is purely technical as a representation of the \enquote{square root} of the density field and cannot be used as an actual \enquote{wavefunction}, since there is no phase information.%
}

\section{Setting the overdensity criterion}
Choosing a sensible value for the overdensity criterion is essential in devising an accurate way to compare the halo catalogues from the grid-based and FOF methods. In a traditional CDM halo analysis there are a few different overdensity choices. The definition of a spherical overdensity halo often uses an average overdensity value of $\rho/\bar{\rho} = 200$, whereas the definitions of virial overdensity may use values in the range of $\rho/\bar{\rho} = \numrange{100}{180}$, depending on the matter density parameter. In a traditional FOF halo finder, the linking length defines an approximate local isodensity contour rather than a mean enclosed overdensity. For a linking length of $b=0.2$ times the mean inter-particle separation, the corresponding local boundary density is approximately $\rho/\bar{\rho} = 80$ \citep{More2011}. The mean enclosed overdensity of the resulting haloes is closer to the usual virial value, $\sim180$ relative to the mean matter density. Our grid-based halo finder also applies a local density threshold, so its overdensity parameter is more directly comparable to the FOF boundary density than to spherical-overdensity definitions. We therefore adopt a value of $\rho/\bar{\rho} = 60$ , which lies in the range expected for the local density contour associated with standard FOF haloes for $b=0.2$. This is the default value for the grid-based halo finder for the value of the matter density parameter adopted in the simulations. This value of $\rho/\bar{\rho}$ is in a similar range to the thresholds for FOF and spherical overdensity methods which allows for meaningful comparison between methods, but as FDM smooths the density field and makes the halo edges less sharp, having a lower overdensity threshold is physically sensible. The grid-based finder operates on a continuous density field, so its threshold does not map directly onto that of particle-based halo finders. We therefore adopt $\rho/\bar{\rho}=60$ consistently throughout the main analysis.


\bsp	
\label{lastpage}
\end{document}